\documentclass[12pt]{article}
\usepackage[latin9]{inputenc}
\usepackage{xcolor}
\usepackage{amsmath}
\usepackage{amsthm}
\usepackage{amssymb}
\usepackage{graphicx}
\usepackage{setspace}
\usepackage[authoryear]{natbib}
\usepackage[unicode=true]
 {hyperref}

\makeatletter
\theoremstyle{plain}
\newtheorem{thm}{\protect\theoremname}
\theoremstyle{definition}
 \newtheorem{example}{\protect\examplename}

\@ifundefined{date}{}{\date{}}
\usepackage{url}
\hypersetup{colorlinks=true, linkcolor=black, citecolor=black, urlcolor=black}
\newcommand{\blind}{0}

\usepackage{changepage}

\newtheorem{proofTheorem}{Proof of Theorem}

\@ifundefined{showcaptionsetup}{}{%
 \PassOptionsToPackage{caption=false}{subfig}}
\usepackage{subfig}
\makeatother

\providecommand{\examplename}{Example}
\providecommand{\theoremname}{Theorem}

\begin{document}
\global\long\def\spacingset{ \global\long
\global\long\def\baselinestretch{%
}%
\small\normalsize}%
\if0\blind { 
\title{\textbf{Effect Decomposition of Functional-Output Computer Experiments
via Orthogonal Additive Gaussian Processes }}
\author{\noindent Yu Tan, Yongxiang Li\thanks{Contact: Yongxiang Li, \protect\href{http://yongxiangli@sjtu.edu.cn}{yongxiangli@sjtu.edu.cn},
Department of Industrial Engineering and Management, Shanghai Jiao
Tong University, Shanghai 200240. } \\
 {\small{}Department of Industrial Engineering and Management, \vspace{-4mm}}\\
{\small{}Shanghai Jiao Tong University}\\
Xiaowu Dai\\
{\small{}Department of Statistics and Data Science and Department
of Biostatistics, \vspace{-4mm}}\\
{\small{}University of California at Los Angeles}\\
Kwok-Leung Tsui\\
{\small{}Department of Industrial, Manufacturing, and Systems Engineering,
\vspace{-4mm}}\\
{\small{}The University of Texas at Arlington}\\
}

\maketitle
} \fi

\noindent \if1\blind { \bigskip{}
\vphantom{} \bigskip{}
\vphantom{} \bigskip{}
\textbf{\LARGE{} }{\LARGE\par}
\noindent \begin{center}
\textbf{\LARGE{}Effect Decomposition of Functional-Output Computer
Experiments via Orthogonal Additive Gaussian Processes}{\LARGE\par}
\par\end{center}

\begin{center}
\par\end{center}

\medskip{}
 } \fi 
\begin{abstract}
\begin{spacing}{1}Functional ANOVA (FANOVA) is a widely used variance-based
sensitivity analysis tool. However, studies on functional-output FANOVA
remain relatively scarce, especially for black-box computer experiments,
which often involve complex and nonlinear functional-output relationships
with unknown data distribution. Conventional approaches often rely
on predefined basis functions or parametric structures that lack the
flexibility to capture complex nonlinear relationships. Additionally,
strong assumptions about the underlying data distributions further
limit their ability to achieve a data-driven orthogonal effect decomposition.
To address these challenges, this study proposes a functional-output
orthogonal additive Gaussian process (FOAGP) to efficiently perform
the data-driven orthogonal effect decomposition. By enforcing a conditional
orthogonality constraint on the separable prior process, the proposed
functional-output orthogonal additive kernel enables data-driven orthogonality
without requiring prior distributional assumptions. The FOAGP framework
also provides analytical formulations for local Sobol' indices and
expected conditional variance sensitivity indices, enabling comprehensive
sensitivity analysis by capturing both global and local effect significance.
Validation through two simulation studies and a real case study on
fuselage shape control confirms the model's effectiveness in orthogonal
effect decomposition and variance decomposition, demonstrating its
practical value in engineering applications. \end{spacing}
\end{abstract}
\begin{spacing}{1}\textit{Keywords:} Computer experiments; Surrogate
modeling; Kriging; Functional-output FANOVA; Effect decomposition.
\end{spacing}

\section{Introduction \label{sec:intro}}

\sloppy

Functional ANOVA (FANOVA, \citealt{efronJackknifeEstimateVariance1981,sobolGlobalSensitivityIndices2001})
has become one of the most popular variance-based sensitivity analyses
\citep{zhangSensitivityAnalysisWind2024}, widely applied in ischemic
heart screening \citep{zhangNewTestFunctional2019}, additive manufacturing
\citep{centofantiRobustFunctionalANOVA2023} and automated machine
learning hyper-parameters tuning \citep{garouaniAutomatedMachineLearning2024}.
By decomposing the output variance and quantifying the importance
of input variables through using Sobol' indices \citep{sobol1990sensitivity,sobol1993sensitivity},
FANOVA serves as a critical tool for variable importance selection
\citep{fang2005design,santnerDesignAnalysisComputer2018}, dimensionality
reduction \citep{muehlenstaedtDatadrivenKrigingModels2012,sungMultiresolutionFunctionalANOVA2020,luAdditiveGaussianProcesses2022a},
and improving model interpretability \citep{chenAlgorithmsEstimateShapley2023}. 

Despite extensive research on FANOVA \citep{muehlenstaedtDatadrivenKrigingModels2012,janonAsymptoticNormalityEfficiency2014,ginsbourgerANOVADecompositionsKernels2016,liuGeneralizedSensitivityAnalysis2021,centofantiRobustFunctionalANOVA2023,marnissiUnifiedViewFANOVA2024},
studies on functional-output FANOVA (FOFANOVA) remain relatively scarce,
especially in the context of computer experiments with black-box functional
outputs (responses), where physical experiments are prohibitively
expensive or computer simulations are time-consuming \citep{kokoszkaIntroductionFunctionalData2017}.
Within the FOFANOVA framework, we study both effect decomposition
and variance decomposition in this study because effect decomposition
is as critical as variance decomposition, with the two being inherently
interconnected and complementary. Specifically, effect decomposition
establishes the theoretical foundation for variance decomposition
\citep{durrandeANOVAKernelsRKHS2013,xiaoMultivariateGlobalSensitivity2018,zhangSensitivityAnalysisWind2024},
while variance decomposition serves as a practical tool to quantify
the significance of the effects and their corresponding input variables
\citep{hookerDiscoveringAdditiveStructure2004,muehlenstaedtDatadrivenKrigingModels2012,sungMultiresolutionFunctionalANOVA2020,apleyVisualizingEffectsPredictor2020}.
Moreover, effect decomposition provides insights into input-output
relationships in black-box computer experiments, thereby improving
model interpretability \citep{duvenaudAdditiveGaussianProcesses2011,durrandeAdditiveCovarianceKernels2012,sungMultiresolutionFunctionalANOVA2020}.

The first step in applying FOFANOVA to computer experiments is functional-output
surrogate modeling. The general functional-output regression model
dates back to \citet{farawayRegressionAnalysisFunctional1997}, which
is a linear and parametric regression approach. To exploit spatial-temporal
correlation in computer experiments with functional outputs, \citet{shiGaussianProcessFunctional2007}
proposed a nonlinear and nonparametric approach, called Gaussian process
(GP) functional regression. \citet{rougierEfficientEmulatorsMultivariate2008}
tackled the challenge of high-dimensional functional outputs by utilizing
a separable covariance structure, enabling efficient emulation of
functional outputs. To further enhance scalability, \citet{wangMultiFidelityHighOrderGaussian2021}
proposed a multi-fidelity high-order GP by leveraging basis decomposition
to reduce model parameters, which which captures complex correlations
between outputs and integrates multi-fidelity. \citet{maComputerModelEmulation2022}
employed nearest-neighbor GP with built-in input dimensionality reduction
via active subspaces, which significantly reduced the dimensionality
of both input and output spaces. To adequately capture output spatial
correlations, \citet{liuLatentFunctionalGaussian2024} proposed a
novel one-stage method called latent functional GP. 

The primary application of FOFANOVA is to identify important input
variables through variance decomposition. To assess the importance
of each input on the time-series output, \citet{drigneiFunctionalANOVAComputer2010}
applied FANOVA at each time point and averaged the FANOVA results
to obtain the expected conditional variance (ECV) sensitivity indices.
\citet{shiCrosscovarianceBasedGlobal2018} extended FANOVA by incorporating
temporal covariance and applying covariance decomposition to derive
a global dynamic sensitivity index, which measures the variable importance
at different time points. It has been shown that that while many computer
experiments have complex functional outputs, these functional outputs
are often driven by a relatively small subset of important input variables
\citep{drigneiFunctionalANOVAComputer2010,wagenerWhatHasGlobal2019,jiangMultipletargetRobustDesign2021,chiognaSensitivityAnalysisWavelet2024}. 

Although these methods perform well in variance decomposition within
FOFANOVA, relatively little attention has been paid to the effect
decomposition of functional outputs or functional-output effect decomposition
(FOED).\textcolor{orange}{{} }Existing methods \citep{chengTimevariantReliabilityAnalysis2019,yueBayesianGeneralizedTwoway2019,bolinFunctionalANOVAModelling2021}
for FOED largely rely on linear or parametric functional regression,
and the use of predefined (user-specified) basis functions imposes
additional assumptions on the underlying data distribution for orthogonality,
such as uniformity. In addition, the lack of data-driven orthogonality
can lead to incorrect attributions of variable/effect importance,
as demonstrated in the case of scalar outputs \citep{durrandeAdditiveCovarianceKernels2012,luAdditiveGaussianProcesses2022a}.
These limitations can lead to biased or inaccurate regression for
datasets from black-box computer experiments, which often involve
complex and nonlinear relationships with unknown data distribution
\citep{santnerDesignAnalysisComputer2018}. The nonlinearity and the
unknown data distribution in functional-output computer experiments
\citep{hookerGeneralizedFunctionalANOVA2007,raoModernNonlinearFunctiononfunction2023,centofantiRobustFunctionalANOVA2023}
pose significant challenges for FOFANOVA.

To address the aforementioned issues, this study proposes a data-driven
methodology for efficient FOED, which is referred to as functional-output
orthogonal additive Gaussian process (FOAGP). Initially, functional
outputs are modeled by a sum of several effects, including high-order
effects. To enforce orthogonality on the underlying data distribution,
the functional-output orthogonal additive kernel is proposed, which
enables data-driven orthogonality and transforms each effect into
an orthogonal one. By incorporating the proposed kernel into the AGP
framework, a nonlinear and nonparametric FOAGP model is proposed for
FOED. Furthermore, the design of the proposed kernel utilizes the
Hadamard product to ensure that the computational complexity of the
FOAGP modeling remains comparable to the standard GP modeling. We
then prove that the FOED based on the proposed FOAGP model is exactly
the orthogonal decomposition. 

Based on the proposed FOED, the corresponding variance decomposition
is further established for sensitivity analysis. Within the FOAGP
framework, we derive analytical local Sobol' indices for local sensitivity
analysis, which capture the local effect significance at each output
position. Additionally, we derive analytical ECV sensitivity indices
for global sensitivity analysis, which measure the whole contribution
of effects over the whole output space. For practical purposes, the
corresponding estimators for both sensitivity analyses are provided.
It is worth noting that all the estimators for the sensitivity analysis
rely solely on the training dataset, enabling global local sensitivity
analysis without prior knowledge about the underlying data distribution.
The proposed FOED and global local sensitivity analysis make the proposed
FOAGP model a unified data-driven implementation of FOFANOVA.  

The remainder of the paper is organized as follows. Sec. \ref{sec:Review}
provides a brief review of FANOVA for computer experiments with scalar
and time-series outputs and orthogonal additive GP (OAGP). Sec. \ref{sec:FOAGP}
introduces the proposed FOAGP framework, covering the modeling, FOED,
and global local variance decomposition. Sec. \ref{sec:Numerical-Examples}
provides simulation examples to demonstrate the effectiveness of FOAGP.
A real case study on fuselage shape control in Sec. \ref{sec:Real-Case-Study}
further validates the the applicability and effectiveness of FOAGP.
Finally, Sec. \ref{sec:Conclusions} concludes the paper and outlines
possible directions for future research. All the proofs of the theorems
in this study can be found in the appendix.

\section{Review of Previous Work \label{sec:Review} }

This section begins with a brief review of FANOVA for computer experiments
with scalar and time-series outputs, followed by a discussion of the
time-variant high dimensional model representation. It then provides
a summary of OAGP and its connection to FANOVA.

\subsection{FANOVA}

FANOVA is a sophisticated tool for global sensitivity analysis in
computer experiments, allowing analysts to decompose the output of
a computer model into its fundamental effects and quantify the significance
of each effect \citep{owenSobolIndicesShapley2014}. Let $f:\mathbb{R}^{d}\rightarrow\mathbb{R}$
be a square-integrable function of $\boldsymbol{x}=\left(x_{i}\right)_{i\in\mathcal{D}}$
and the components $\left\{ x_{i}\right\} _{i\in\mathcal{D}}$ are
assumed to be independent, where $\mathcal{D}=\left\{ 1,\cdots,d\right\} $.
The FANOVA decomposition \citep{sobol1990sensitivity,owenShapleyValueMeasuring2017}
of $f\left(\boldsymbol{x}\right)$ is defined as 
\begin{align}
f\left(\boldsymbol{x}\right)= & f_{0}+f_{1}\left(x_{1}\right)+f_{2}\left(x_{2}\right)+\cdots+f_{12}\left(x_{1},x_{2}\right)+\cdots+f_{12\cdots d}\left(x_{1},\cdots,x_{d}\right)=\sum_{\boldsymbol{u}\subseteq\mathcal{D}}f_{\boldsymbol{u}}\left(\boldsymbol{x}_{\boldsymbol{u}}\right),\label{eq:FANOVA}
\end{align}
where $f_{\boldsymbol{u}}\left(\boldsymbol{x}_{\boldsymbol{u}}\right)$
is a function of $\boldsymbol{x}_{\boldsymbol{u}}=\left(x_{i}\right)_{i\in\boldsymbol{u}}$
corresponding to a subset $\boldsymbol{u}$ of $\mathcal{D}$, and
is referred to as the effect function or effect for simplicity. When
$\boldsymbol{u}=\varnothing$, the effect function $f_{\varnothing}$
degenerates to a constant and is denoted as $f_{0}$. The constant
effect $f_{0}$ is defined as the expectation of $f\left(\boldsymbol{x}\right)$,
denoted as $\mathbb{E}_{\boldsymbol{x}}\left[f\left(\boldsymbol{x}\right)\right]$,
and the remaining effects $\left\{ f_{\boldsymbol{u}}\right\} _{\boldsymbol{u}\ne\varnothing}$
are determined by the orthogonality constraints 
\begin{equation}
\int f_{\boldsymbol{u}}\left(\boldsymbol{x}_{\boldsymbol{u}}\right)\mathrm{d}F_{i}\left(x_{i}\right)=0\left(\forall i\in\boldsymbol{u}\right),\label{eq:FANOVA Cons}
\end{equation}
where $F_{i}\left(x_{i}\right)$ is the cumulative cumulative distribution
function (CDF) of the $i$th component $x_{i}$.  Due to the orthogonality
between the effects$\left\{ f_{\boldsymbol{u}}\right\} $, the FANOVA
decomposition in Eq. \eqref{eq:FANOVA} is also referred to as orthogonal
effect decomposition. 

The variance decomposition of FANOVA directly follows the orthogonal
effect decomposition: 
\[
\mathbb{V}_{\boldsymbol{x}}\left(f\left(\boldsymbol{x}\right)\right)=\sum_{\boldsymbol{u}\subseteq\mathcal{D}}\mathbb{V}_{\boldsymbol{x}_{\boldsymbol{u}}}\left(f_{\boldsymbol{u}}\left(\boldsymbol{x}_{\boldsymbol{u}}\right)\right),
\]
where $\mathbb{V}_{\boldsymbol{x}}\left(f\left(\boldsymbol{x}\right)\right)$
denotes the variance of $f\left(\boldsymbol{x}\right)$, i.e., $\mathbb{V}_{\boldsymbol{x}}\left(f\left(\boldsymbol{x}\right)\right)=\mathbb{E}_{\boldsymbol{x}}\left[\left(f\left(\boldsymbol{x}\right)-\mathbb{E}_{\boldsymbol{x}}\left[f\left(\boldsymbol{x}\right)\right]\right)^{2}\right]$.
Based on the variance decomposition, the relative importance of component
subsets $\left\{ \boldsymbol{u}\right\} _{\boldsymbol{u}\subseteq\mathcal{D}}$
can be quantified by Sobol' indices \citep{sobol1990sensitivity,sobol1993sensitivity},
which is defined as 
\begin{equation}
\phi_{\boldsymbol{u}}=\frac{\mathbb{V}_{\boldsymbol{x}_{\boldsymbol{u}}}\left(f_{\boldsymbol{u}}\left(\boldsymbol{x}_{\boldsymbol{u}}\right)\right)}{\mathbb{V}_{\boldsymbol{x}}\left(f\left(\boldsymbol{x}\right)\right)}=\frac{\mathbb{V}_{\boldsymbol{x}_{\boldsymbol{u}}}\left(f_{\boldsymbol{u}}\left(\boldsymbol{x}_{\boldsymbol{u}}\right)\right)}{\sum_{\boldsymbol{v}\subseteq\mathcal{D}}\mathbb{V}_{\boldsymbol{x}_{\boldsymbol{v}}}\left(f_{\boldsymbol{v}}\left(\boldsymbol{x}_{\boldsymbol{v}}\right)\right)}.\label{eq:Sobol' Indices}
\end{equation}

To address the variance dominance of $t$ in time series and assess
the importance of each input on the time-series outputs, \citet{drigneiFunctionalANOVAComputer2010}
theoretically applied FANOVA at any fixed time point to obtain the
time-conditional orthogonal effect decomposition $f\left(\boldsymbol{x},t\right)=\sum_{\boldsymbol{u}\subseteq\mathcal{D}}f_{\boldsymbol{u}\mid t}\left(\boldsymbol{x}_{\boldsymbol{u}},t\right)$
and took the expectation of time-conditional variance with respect
to time $t$ to derive the ECV decomposition $\text{\ensuremath{\mathbb{E}_{t}\left[\mathbb{V}_{\boldsymbol{x}}\left(f\left(\boldsymbol{x},t\right)\right)\right]}}=\sum_{\boldsymbol{u}\subseteq\mathcal{D}}\mathbb{E}_{t}\left[\mathbb{V}_{\boldsymbol{x}_{\boldsymbol{u}}}\left(f_{\boldsymbol{u}\mid t}\left(\boldsymbol{x}_{\boldsymbol{u}},t\right)\right)\right]$.
Normalizing the ECV $\mathbb{E}_{t}\left[\mathbb{V}_{\boldsymbol{x}_{\boldsymbol{u}}}\left(f_{\boldsymbol{u}\mid t}\left(\boldsymbol{x}_{\boldsymbol{u}},t\right)\right)\right]$
yielded the ECV sensitivity indices  
\begin{align}
\mathcal{S}_{\boldsymbol{u}} & =\frac{\mathbb{E}_{t}\left[\mathbb{V}_{\boldsymbol{x}_{\boldsymbol{u}}}\left(f_{\boldsymbol{u}\mid t}\left(\boldsymbol{x}_{\boldsymbol{u}},t\right)\right)\right]}{\sum_{\boldsymbol{v}\subseteq\mathcal{D}}\mathbb{E}_{t}\left[\mathbb{V}_{\boldsymbol{x}_{\boldsymbol{v}}}\left(f_{\boldsymbol{v}\mid t}\left(\boldsymbol{x}_{\boldsymbol{v}},t\right)\right)\right]}.\label{eq:ECV Indices (Review)}
\end{align}
To circumvent the time-conditional orthogonal effect decomposition
of time-series outputs, \citet{drigneiFunctionalANOVAComputer2010}
employed the standard orthogonal effect decomposition incorporating
$t$ as an additional variable to calculate the ECV sensitivity indices
through
\begin{equation}
\mathcal{S}_{\boldsymbol{u}}=\phi_{\boldsymbol{u}}+\phi_{\boldsymbol{u},t},\label{eq:ECV Indices based on FANOVA}
\end{equation}
where numerical integration techniques were used. 

\citet{chengTimevariantReliabilityAnalysis2019} introduced the time-variant
high dimensional model representation (HDMR) $f\left(\boldsymbol{x},t\right)\approx\sum_{\left|\boldsymbol{u}\right|\leqslant2}f_{\boldsymbol{u}\mid t}\left(\boldsymbol{x}_{\boldsymbol{u}},t\right)$,
where $\left|\boldsymbol{u}\right|$ is the cardinality of the set
$\boldsymbol{u}$, and further approximated the effect functions using
a set of orthogonal polynomials, yielding

\begin{align}
f\left(\boldsymbol{x},t\right)\approx & \sum_{i=1}^{o}\alpha_{i}\varphi_{i}\left(t\right)+\sum_{i=1}^{d}\sum_{p+q\leqslant o}\beta_{pq}^{\left(i\right)}\varphi_{p}\left(x_{i}\right)\varphi_{q}\left(t\right)\nonumber \\
 & +\sum_{1\leqslant i<j\leqslant d}\sum_{p+q+r\leqslant o}\gamma_{pqr}^{\left(ij\right)}\varphi_{p}\left(x_{i}\right)\varphi_{q}\left(x_{j}\right)\varphi_{r}\left(t\right),\label{eq:HDMR}
\end{align}
where $\left\{ \varphi_{p}\left(\cdot\right)\right\} _{p=1,2,\cdots}$
denoted the orthogonal polynomials basis functions of order $p$,
$\left\{ \alpha_{i},\beta_{pq}^{\left(i\right)},\gamma_{pqr}^{\left(ij\right)}\right\} $
represented the basis function coefficients and $o$ was the truncated
order of the polynomial basis functions. The coefficients $\left\{ \alpha_{i},\beta_{pq}^{\left(i\right)},\gamma_{pqr}^{\left(ij\right)}\right\} $
can be estimated within a Gaussian process framework \citep{chengTimevariantReliabilityAnalysis2019}.
For simplicity, we will refer to the truncated HDMR model in Eq.
\eqref{eq:HDMR} as the HDMR model throughout the rest of the study. 

When the input variable $x_{i}$ follows a uniform distribution on
a specified interval, which is determined by the predefined orthogonal
basis functions, the HDMR model in Eq. \eqref{eq:HDMR} performs the
two-way orthogonal effect decomposition, provided the truncated order
is sufficiently large. However, this reliance on a uniform distribution
introduces a significant limitation in the black-box computer experiments
with functional outputs, where prior knowledge about the underlying
data distribution is often unavailable. Consequently, the orthogonality
assumed by the HDMR model may not hold in practice. Moreover, the
parametric structure and truncated order limits the number of basis
functions, restricting regression flexibility of the HDMR model. This
limitation may lead to biased or inaccurate predictions \citep{sungMultiresolutionFunctionalANOVA2020},
especially when the high-order effects are present in the experiments. 

\subsection{Orthogonal Additive Gaussian Processes }

\textcolor{gray}{}\citet{duvenaudAdditiveGaussianProcesses2011}
proposed AGP by incorporating an additive structure within GP
\begin{align}
f\left(\boldsymbol{x}\right)= & f_{1}\left(x_{1}\right)+f_{2}\left(x_{2}\right)+\cdots+f_{12}\left(x_{1},x_{2}\right)+\cdots+f_{12\cdots d}\left(x_{1},\cdots,x_{d}\right)=\sum_{\boldsymbol{u}\subseteq\mathcal{D}}f_{\boldsymbol{u}}\left(\boldsymbol{x}_{\boldsymbol{u}}\right),\label{eq:AGP}
\end{align}
where each effect function $f_{\boldsymbol{u}}\left(\boldsymbol{x}_{\boldsymbol{u}}\right)$
is also a GP. This additive structure is the same as that of orthogonal
effect decomposition in Eq. \eqref{eq:FANOVA}, except for the absence
of the constant effect $f_{0}$. In practice, the additive structure
is enforced through the use of an additive kernel. Given the kernel
$k_{i}\left(\cdot\right)$ associated with the $i$th input $x_{i}$,
the $n$th-order additive kernel \citep{duvenaudAdditiveGaussianProcesses2011}
is defined as 
\begin{equation}
k_{add_{n}}\left(\boldsymbol{x},\boldsymbol{x}^{\prime}\right)=\lambda_{n}^{2}\sum_{\left|\boldsymbol{u}\right|=n}\prod_{i\in\boldsymbol{u}}k_{i}\left(x_{i},x_{i}^{\prime}\right)\label{eq:Hierarchical Kernels}
\end{equation}
for $1\leqslant n\leqslant d$, where $\lambda_{n}^{2}$ is the $n$th-order
variance, and $x_{i},x_{i}^{\prime}$ are the $i$th components of
$\boldsymbol{x},\boldsymbol{x}^{\prime}$, respectively. 

\citet{luAdditiveGaussianProcesses2022a} proposed OAGP by applying
the orthogonality constraints in Eq. \eqref{eq:FANOVA Cons} on effect
functions $\left\{ f_{\boldsymbol{u}}\left(\boldsymbol{x}_{\boldsymbol{u}}\right)\right\} $
in Eq. \eqref{eq:AGP}, resulting in an orthogonal additive kernel
for each effect function. To capture the mean and ensure identifiability,
the constant mean effect $f_{0}$ is also included. Using this construction,
\citet{luAdditiveGaussianProcesses2022a} demonstrated that OAGP performs
exact orthogonal effect decomposition of scalar outputs, and deduced
the analytical formulas of Sobol' indices for practical variance decomposition. 

\section{Functional-Output Orthogonal Additive Gaussian Process \label{sec:FOAGP} }

This section details the proposed FOAGP framework, beginning with
a description of its additive structure in Sec. \ref{subsec:Model-Formulation},
followed by the proposed functional-output orthogonal additive kernel
in Sec. \ref{subsec:OATS Kernel}. The parameter estimation of FOAGP
is provided in Sec. \ref{subsec:Parameter-Estimation}. Sec. \ref{subsec:FO Effect Decom}
reveals that the proposed FOAGP model performs exactly orthogonal
decomposition. Finally, Sec. \ref{subsec:Variance Decomposition}
details the global local variance decomposition within FOAGP framework
and provides analytical local Sobol' indices and\textcolor{black}{{}
analytical ECV sensitivity indices}, enabling comprehensive sensitivity
analysis. 

\subsection{Model Formulation of FOAGP \label{subsec:Model-Formulation}}

In this study, we only consider one-dimensional functional output
case for simplicity. In fact, many computer experiments, such as the
fuselage shape control \citet{liuLatentFunctionalGaussian2024}, only
have one-dimensional functional outputs. 

Denote the response of the input $\boldsymbol{x}$ at the output position
$t$ by 
\begin{equation}
y\left(\boldsymbol{x},t\right)=f\left(\boldsymbol{x},t\right)+\epsilon,\label{eq:Model0}
\end{equation}
where $f\left(\boldsymbol{x},t\right)$ follows an AGP, and $\epsilon\sim\mathcal{N}\left(0,\sigma^{2}\delta_{0}^{2}\right)$
represents the independent and identically distributed measurement
errors that are independent of $f\left(\boldsymbol{x},t\right)$.
In this study, we assume that the proposed model has the following
additive structure 
\begin{align}
f\left(\boldsymbol{x},t\right) & =\sum_{\boldsymbol{u}\subseteq\mathcal{D}}f_{\boldsymbol{u}\mid t}\left(\boldsymbol{x}_{\boldsymbol{u}},t\right)\nonumber \\
 & =f_{0\mid t}\left(t\right)+f_{1\mid t}\left(x_{1},t\right)+\cdots+f_{12\mid t}\left(x_{1},x_{2},t\right)+\cdots+f_{1\cdots d\mid t}\left(x_{1},\cdots,x_{d},t\right),\label{eq:FOAGP}
\end{align}
where each functional-output effect function $f_{\boldsymbol{u}\mid t}\left(\boldsymbol{x}_{\boldsymbol{u}},t\right)$
also follows a GP with zero mean. For simplicity, $f_{\boldsymbol{u}\mid t}\left(\boldsymbol{x}_{\boldsymbol{u}},t\right)$
will be referred to as the functional-output effect or simply the
effect throughout this study. The decomposition of $f\left(\boldsymbol{x},t\right)$
in Eq. \eqref{eq:FOAGP} is called FOED. 

Unlike the OAGP model with a global constant mean effect \citep{luAdditiveGaussianProcesses2022a},
the proposed model includes a mean effect $f_{0\mid t}\left(t\right)$
that depends on the output position $t$ and is determined by a conventional
kernel $k_{t}\left(\cdot\right)$ without orthogonality constraints
\citep{durrandeAdditiveCovarianceKernels2012}. This mean effect $f_{0\mid t}\left(t\right)$
extracts the mean of the functional output $y\left(\boldsymbol{x},t\right)$
at any output position, enabling other effects $\left\{ f_{\boldsymbol{u}\mid t}\left(\boldsymbol{x}_{\boldsymbol{u}},t\right)\mid\boldsymbol{u}\ne\varnothing\right\} $
to have zero mean. Without this mean effect $f_{0\mid t}\left(t\right)$,
the model may have identifiability issues because any constant could
be added to one effect and subtracted from another. 

Instead of directly applying the traditional FANOVA decomposition
as in Eq. \eqref{eq:FANOVA} by treating the output position $t$
as an additional variable, the proposed model focuses on the effect
decomposition with respect to the input $\boldsymbol{x}$. One key
reason for this is that the variance dominance of $t$ can obscure
the importance of input variables $\boldsymbol{x}$ \citep{drigneiFunctionalANOVAComputer2010}.
Furthermore, the output position $t$ is an intrinsic variab\textcolor{black}{le
and is of no interest from a sensitivity analysis perspective.} With
carefully designed kernels in Sec. \ref{subsec:OATS Kernel}, the
effects $\left\{ f_{\boldsymbol{u}\mid t}\left(\boldsymbol{x}_{\boldsymbol{u}},t\right)\right\} $
are conditionally orthogonal in $L_{2}$ at any output position $t$,
which ensures model identifiability. Because of these properties above,
the proposed model $f\left(\boldsymbol{x},t\right)$ in Eq. \eqref{eq:FOAGP}
is called the FOAGP model. 

\subsection{Functional-Output Orthogonal Additive Kernel \label{subsec:OATS Kernel}}

To achieve the orthogonality between effects $\left\{ f_{\boldsymbol{u}\mid t}\left(\boldsymbol{x}_{\boldsymbol{u}},t\right)\right\} $,
specific constraints must be imposed. First, we propose a separable
prior process for the effect $f_{\boldsymbol{u}\mid t}\left(\boldsymbol{x}_{\boldsymbol{u}},t\right)$,
which is determined by the following kernel 
\begin{equation}
k_{\boldsymbol{u}\mid t}\left(\left(\boldsymbol{x},t\right),\left(\boldsymbol{x}^{\prime},t^{\prime}\right)\right)=k_{t}\left(t,t^{\prime}\right)\prod_{i\in\boldsymbol{u}}k_{i}\left(x_{i},x_{i}^{\prime}\right),\label{eq:Sep Kernel}
\end{equation}
where $k_{i}\left(\cdot\right)$ is the radial kernel of the $i$th
input $x_{i}$ with an unknown parameter $\theta_{i}$, $k_{t}\left(\cdot\right)$
is the radial kernel of output position $t$ with an unknown parameter
$\theta_{t}$. The vector of these parameters are denoted by $\boldsymbol{\theta}=\left[\theta_{1},\cdots\theta_{d},\theta_{t}\right]^{T}$.
In this study, $k_{i}\left(\cdot\right)$ and $k_{t}\left(\cdot\right)$
are called the input and output kernels, respectively. The kernel
structure of the separable prior process in Eq. \eqref{eq:Sep Kernel}
assumes that there is no input-output interaction because the input
and output kernels share no common variables, which is reasonable
in many applications \citep{todescatoEfficientSpatiotemporalGaussian2020,hamelijnckSpatioTemporalVariationalGaussian2021,lambardidisanminiatoSeparableSpatiotemporalKriging2022,zhangEfficientImplementationSpatial2023}. 

As the output position $t$ is not involved into the effect decomposition,
the orthogonality constraints as provided in Eq. \eqref{eq:FANOVA Cons}
cannot be directly applied to the functional-output effects $\left\{ f_{\boldsymbol{u}\mid t}\left(\boldsymbol{x}_{\boldsymbol{u}},t\right)\right\} $.
Instead, consider the conditional orthogonality constraint 
\begin{equation}
\int f_{\boldsymbol{u}\mid t}\left(\boldsymbol{x}_{\boldsymbol{u}},t\right)\mathrm{d}F_{i}\left(x_{i}\right)=0\label{eq:C-OCons}
\end{equation}
for any $i\in\boldsymbol{u}$ and output position $t$. We first determine
the conditional process $f_{i\mid t}\mid\left\{ \int f_{i\mid t}\left(x_{i},t\right)\mathrm{d}F_{i}\left(x_{i}\right)=0\right\} $
of the main effect $f_{i\mid t}\left(x_{i},t\right)$, followed by
the conditional process $f_{\boldsymbol{u}\mid t}\mid\left\{ \int f_{\boldsymbol{u}\mid t}\left(\boldsymbol{x}_{\boldsymbol{u}},t\right)\mathrm{d}F_{i}\left(x_{i}\right)=0\right\} $
of the interaction effect $f_{\boldsymbol{u}\mid t}\left(\boldsymbol{x}_{\boldsymbol{u}},t\right)$.
To determine\textcolor{teal}{{} }the first-order conditional process
\begin{equation}
f_{i\mid t}\mid\left\{ \int f_{i\mid t}\left(x_{i},t\right)\mathrm{d}F_{i}\left(x_{i}\right)=0\right\} ,\label{eq:Cond_Process}
\end{equation}
the following theorem is proposed. 
\begin{thm}
\label{thm:Cond GP} Suppose the separable prior process $f_{i\mid t}$
is a GP determined by the kernel $k_{i}\left(\cdot\right)k_{t}\left(\cdot\right)$.
Under the assumption that $x_{i}$ and $t$ are independent, the conditional
process $f_{i\mid t}\mid\left\{ \int f_{i\mid t}\left(x_{i},t\right)\mathrm{d}F_{i}\left(x_{i}\right)=0\right\} $
is also a GP determined by a re-constructed kernel $\tilde{k}_{i}\left(\cdot\right)k_{t}\left(\cdot\right)$,
where 
\begin{equation}
\tilde{k}_{i}\left(x_{i},x_{i}^{\prime}\right)=k_{i}\left(x_{i},x_{i}^{\prime}\right)-\frac{\mathbb{E}_{x}\left[k_{i}\left(x,x_{i}\right)\right]\mathbb{E}_{x}\left[k_{i}\left(x,x_{i}^{\prime}\right)\right]}{\mathbb{E}_{x,x^{\prime}}\left[k_{i}\left(x,x^{\prime}\right)\right]}.\label{eq:OAKernel}
\end{equation}
\end{thm}
Theorem \ref{thm:Cond GP} not only establishes that the conditional
process follows a GP, but also provides the analytical formula of
the re-constructed kernel. The re-constructed kernels $\left\{ \tilde{k}_{i}\left(\cdot\right)\right\} $
are called orthogonal kernels. These orthogonal kernels $\left\{ \tilde{k}_{i}\left(\cdot\right)\right\} $
are identical to those obtained by \citet{durrandeANOVAKernelsRKHS2013,plumleeOrthogonalGaussianProcess2018}
due to the assumption of no input-output interactions. Moreover, previous
proofs for computer experiments with scalar outputs often rely on
reproducing kernel Hilbert spaces or stochastic processes, which can
be less intuitive and may not be directly extended to functional-output
cases. In contrast, this study provides a proof for functional-output
computer experiments based on the posterior distribution of the GP,
offering a more accessible and practical approach. In practice, the
expectations $\mathbb{E}_{x}\left[k_{i}\left(x,x_{i}\right)\right]$
and $\mathbb{E}_{x,x'}\left[k_{i}\left(x,x^{\prime}\right)\right]$
can be estimated using the training data as an empirical distribution
\citep{luAdditiveGaussianProcesses2022a}. 

It can be shown that the the orthogonal kernels $\left\{ \tilde{k}_{i}\left(\cdot\right)\right\} $
are positive semi-definite \citep{plumleeOrthogonalGaussianProcess2018}.
Moreover, these orthogonal kernels $\left\{ \tilde{k}_{i}\left(\cdot\right)\right\} $
are data-driven and nonlinear, allowing them to capture the underlying
data distribution and enable accurate regression. Furthermore, these
orthogonal kernels $\left\{ \tilde{k}_{i}\left(\cdot\right)\right\} $
are generally non-isotropic, non-stationary and can take both positive
and negative values, enhancing their flexibility in modeling intricate
relationships \citep{luAdditiveGaussianProcesses2022a}.  

Theorem \ref{thm:Cond GP} can be extended to the high-order conditional
process $f_{\boldsymbol{u}\mid t}\mid\left\{ \int f_{\boldsymbol{u}\mid t}\left(\boldsymbol{x}_{\boldsymbol{u}},t\right)\mathrm{d}F_{i}\left(x_{i}\right)=0,\forall i\in\boldsymbol{u}\right\} $
of the interaction effect $f_{\boldsymbol{u}\mid t}$ with minor modifications.
This extension leverages the proposed separable prior process, which
simplifies the enforcement of the conditional orthogonality constraint.
 As a result, the conditional process $f_{\boldsymbol{u}\mid t}\mid\left\{ \int f_{\boldsymbol{u}\mid t}\left(\boldsymbol{x}_{\boldsymbol{u}},t\right)\mathrm{d}F_{i}\left(x_{i}\right)=0,\forall i\in\boldsymbol{u}\right\} $
of the interaction effect $f_{\boldsymbol{u}\mid t}$ also follows
a GP determined by a re-constructed kernel $\tilde{k}_{\boldsymbol{u}\mid t}\left(\cdot\right)$,
where 
\begin{equation}
\tilde{k}_{\boldsymbol{u}\mid t}\left(\left(\boldsymbol{x}_{\boldsymbol{u}},t\right),\left(\boldsymbol{x}_{\boldsymbol{u}}^{\prime},t^{\prime}\right)\right)=k_{t}\left(t,t^{\prime}\right)\prod_{i\in\boldsymbol{u}}\tilde{k}_{i}\left(x_{i},x_{i}^{\prime}\right).\label{eq:FOA Kernel}
\end{equation}
It is worth noting that the re-constructed kernel $\tilde{k}_{i\mid t}\left(\cdot\right)=\tilde{k}_{i}\left(\cdot\right)k_{t}\left(\cdot\right)$
in Theorem \ref{thm:Cond GP} also follows the structure in Eq. \eqref{eq:FOA Kernel}.
The proposed kernels $\left\{ \tilde{k}_{\boldsymbol{u}\mid t}\left(\cdot\right)\right\} $
are called functional-output orthogonal additive kernels. For the
remainder of this study, the notation $f_{\boldsymbol{u}\mid t}\left(\boldsymbol{x}_{\boldsymbol{u}},t\right)$
is used to denote the conditional process $f_{\boldsymbol{u}\mid t}\mid\left\{ \int f_{\boldsymbol{u}\mid t}\left(\boldsymbol{x}_{\boldsymbol{u}},t\right)\mathrm{d}F_{i}\left(x_{i}\right)=0\right\} $
for simplicity. It can be shown that the proposed kernels enable all
the effects $\left\{ f_{\boldsymbol{u}\mid t}\right\} $ to satisfy
the conditional orthogonality constraints in Eq. \eqref{eq:C-OCons}.
Since each pair of the effects contains at least one non-common variable,
it can be shown that the conditional GPs are conditionally orthogonal
in $L_{2}$ space, which guarantees the identifiability of the proposed
FOAGP model. 

\subsection{Parameter Estimation \label{subsec:Parameter-Estimation}}

Consider a dataset $\left\{ \boldsymbol{X},\text{\ensuremath{\boldsymbol{T}}},\boldsymbol{y}\right\} $
consisting of $N$ samples in a $d$-dimensional space, where $\boldsymbol{X}=\left[\boldsymbol{x}_{1},\cdots,\boldsymbol{x}_{N}\right]^{\top}=\left(x_{uv}\right)_{N\times d}$
is the input point, $\boldsymbol{T}=\left[t_{1},\cdots,t_{N}\right]^{\top}$
is the measured output position, and $\boldsymbol{y}=\left[y_{1},\cdots,y_{N}\right]^{\top}$
is the corresponding output. Let $\boldsymbol{X}_{i}=\left[x_{1i},\cdots,x_{Ni}\right]^{\top}$
denote the $i$th column of $\boldsymbol{X}$. The input kernel matrix
of $\boldsymbol{X}_{i}$ and output kernel matrix are denoted by $\boldsymbol{K}_{i}=\tilde{k}_{i}\left(\boldsymbol{X}_{i},\boldsymbol{X}_{i}\right)=\left[\tilde{k}_{i}\left(x_{ui},x_{vi}\right)\right]_{u,v\in\left\{ 1,\cdots,N\right\} }$
and $\boldsymbol{K}_{t}=k_{t}\left(\boldsymbol{T},\boldsymbol{T}\right)=\left[k_{t}\left(t_{u},t_{v}\right)\right]_{u,v\in\left\{ 1,\cdots,N\right\} }$,
respectively.\textcolor{orange}{{} }The covariance matrix of FOAGP can
be rewritten into the Hadamard product of $d+1$ matrices as 

\begin{equation}
\sigma^{2}\boldsymbol{K}=\sigma^{2}\left(\delta_{0}^{2}\boldsymbol{I}_{N}+\delta_{t}^{2}\boldsymbol{K}_{t}\odot\left(\bigodot_{i=1}^{d}\left(\mathbf{1}_{N}\mathbf{1}_{N}^{\top}+\delta_{i}^{2}\boldsymbol{K}_{i}\right)\right)\right),\label{eq:Cov Matrix}
\end{equation}
where $\odot$ denotes the Hadamard product, $\boldsymbol{I}_{N}$
is the $N\times N$ identity matrix, $\mathbf{1}_{N}$ is an $N\times1$
vector consisting of $N$ ones, and $\boldsymbol{\delta}=\left[\delta_{0},\delta_{1},\cdots,\delta_{d},\delta_{t}\right]^{\text{\ensuremath{\top}}}$
are parameters to adjust the relative variance across dimensions.
The time complexity of computing $\boldsymbol{K}$ in Eq. \eqref{eq:Cov Matrix}
is $O\left(N^{2}d\right)$, making the FOAGP modeling comparable to
the standard GP modeling. 

Denote the parameters of the kernels by $\boldsymbol{\theta}$ and
let $\boldsymbol{\phi}=\left\{ \boldsymbol{\delta},\boldsymbol{\theta}\right\} $.
Parameter estimation is achieved by maximizing the log marginal likelihood
(up to a constant):

\begin{equation}
\ell\left(\sigma^{2},\boldsymbol{\phi}\right)=-\frac{1}{2}\left(\log\left|\sigma^{2}\boldsymbol{K}\right|+\frac{\boldsymbol{y}^{\top}\boldsymbol{K}^{-1}\boldsymbol{y}}{\sigma^{2}}\right).\label{eq:Likelihood}
\end{equation}
Given $\boldsymbol{\phi}$, maximizing the log-likelihood is equivalent
to solving the equation that the derivative of the log-likelihood
with respect to $\sigma^{2}$ equals to zero. Thus, the parameter
estimator of $\sigma^{2}$ is 
\begin{equation}
\hat{\sigma}^{2}=\frac{1}{N}\boldsymbol{y}^{\top}\boldsymbol{K}^{-1}\boldsymbol{y}.\label{eq:sigma2}
\end{equation}
Finally, the parameter $\boldsymbol{\phi}$\textcolor{green}{{} }can
be numerically optimized by minimizing 
\begin{equation}
\hat{\boldsymbol{\phi}}=\mathrm{arg}\underset{\boldsymbol{\phi}}{\min}\left\{ N\log\hat{\sigma}^{2}+\log\left|\boldsymbol{K}\right|\right\} ,\label{eq:Obj}
\end{equation}
which can be solved with the gradient descent algorithms \citep{lophaven2002aspects}.

 When the functional outputs are collected at grid output position
$\left\{ \tau_{v}\right\} _{v=1}^{n}$ for each input $\boldsymbol{\chi}_{u}\in\mathbb{R}^{d}$,
where $u=1,\cdots,m$ and $N=mn$, the Hadamard product in Eq. \eqref{eq:Cov Matrix}
can be replaced by the Kronecker product. Specifically, the  covariance
matrix can be rewritten as 

\begin{align}
\sigma^{2}\boldsymbol{K} & =\sigma^{2}\left(\delta_{0}^{2}\boldsymbol{I}_{N}+\delta_{t}^{2}\boldsymbol{R}_{t}\otimes\left(\bigodot_{i=1}^{d}\left(\mathbf{1}_{m}\mathbf{1}_{m}^{\top}+\delta_{i}^{2}\boldsymbol{R}_{i}\right)\right)\right)=\sigma^{2}\left(\delta_{0}^{2}\boldsymbol{I}_{N}+\boldsymbol{C}_{t}\otimes\boldsymbol{C}_{\boldsymbol{x}}\right),\label{eq:KronKernel}
\end{align}
where $\otimes$ denotes the Kronecker product, $\boldsymbol{R}_{t}=\left[k_{t}\left(\tau_{u},\tau_{v}\right)\right]_{u,v\in\left\{ 1,\cdots,n\right\} }$,
$\boldsymbol{R}_{i}=\left[\tilde{k}_{i}\left(\chi_{ui},\chi_{vi}\right)\right]{}_{u,v\in\left\{ 1,\cdots,m\right\} }$,
$\boldsymbol{C}_{t}=\delta_{t}^{2}\boldsymbol{R}_{t}$, and $\boldsymbol{C}_{\boldsymbol{x}}=\bigodot_{i=1}^{d}\left(\mathbf{1}_{m}\mathbf{1}_{m}^{\top}+\delta_{i}^{2}\boldsymbol{R}_{i}\right)$.
Applying Eigen decomposition to $\boldsymbol{C}_{t}$ and $\boldsymbol{C}_{\boldsymbol{x}}$
gives $\boldsymbol{C}_{t}=\boldsymbol{U}\boldsymbol{D}\boldsymbol{U}^{\top}$
and $\boldsymbol{C}_{\boldsymbol{x}}=\boldsymbol{V}\boldsymbol{\varLambda}\boldsymbol{V}^{\top}$,
where $\boldsymbol{U}$ and $\boldsymbol{V}$ are orthogonal matrices,
and $\boldsymbol{D}$ and $\boldsymbol{\varLambda}$ are diagonal
matrices containing the eigenvalues of $\boldsymbol{C}_{t}$ and $\boldsymbol{C}_{\boldsymbol{x}}$,
respectively. By applying the Eigen decomposition, the correlation
matrix $\boldsymbol{K}$ can be rewritten as 
\[
\boldsymbol{K}=\left(\boldsymbol{U}\otimes\boldsymbol{V}\right)\left(\delta_{0}^{2}\boldsymbol{I}+\boldsymbol{D}\otimes\boldsymbol{\varLambda}\right)\left(\boldsymbol{U}\otimes\boldsymbol{V}\right)^{\top},
\]
which implies 
\begin{equation}
\begin{cases}
\left|\boldsymbol{K}\right| & =\left|\delta_{0}^{2}\boldsymbol{I}+\boldsymbol{D}\otimes\boldsymbol{\varLambda}\right|=\left|\boldsymbol{S}\right|\\
\boldsymbol{K}^{-1} & =\left(\boldsymbol{U}\otimes\boldsymbol{V}\right)\boldsymbol{S}^{-1}\left(\boldsymbol{U}\otimes\boldsymbol{V}\right)^{\top}
\end{cases},\label{eq:Fast Formula}
\end{equation}
where $\boldsymbol{S}=\delta_{0}^{2}\boldsymbol{I}+\boldsymbol{D}\otimes\boldsymbol{\varLambda}$
is also a diagonal matrix. These results in Eq. \eqref{eq:Fast Formula}
significantly accelerate the computation of $\left|\boldsymbol{K}\right|$
and $\boldsymbol{K}^{-1}$ from $O\left(N^{3}\right)$ to $O\left(m^{3}+n^{3}\right)$.
By plugging Eq. \eqref{eq:Fast Formula} into Eq. \eqref{eq:sigma2},
the parameter estimator of $\sigma^{2}$ becomes
\begin{equation}
\hat{\sigma}^{2}=\frac{1}{N}\left(\mathrm{Vec}\left(\boldsymbol{V}^{\top}\boldsymbol{YU}\right)\right)^{\top}\boldsymbol{S}^{-1}\mathrm{Vec}\left(\boldsymbol{V}^{\top}\boldsymbol{YU}\right).\label{eq:Kron-sigma2}
\end{equation}
In the above equation, $\boldsymbol{Y}$ is the matricization \citep{zhangMatrixAnalysisApplications2017}
of $\boldsymbol{y}$, i.e.,
\[
\boldsymbol{Y}=\left[\begin{matrix}y_{11} & \cdots & y_{1n}\\
\vdots & \ddots & \vdots\\
y_{m1} & \cdots & y_{mn}
\end{matrix}\right],
\]
where $y_{uv}$ is the measurement of $f\left(\boldsymbol{x},t\right)$
of the input $\boldsymbol{\chi}_{u}$ at output position $\tau_{v}$.
$\mathrm{Vec}$ denotes the vectorization that converts a matrix into
a vector by sequentially arranging its columns one after the other
into a single vector \citep{zhangMatrixAnalysisApplications2017}.\textcolor{orange}{{}
}Given the inverse correlation matrix $\boldsymbol{K}^{-1}$, Eq.
\eqref{eq:Kron-sigma2} reduces the time complexity of estimating
$\sigma^{2}$ from $O\left(N^{2}\right)$ of Eq. \eqref{eq:sigma2}
to $O\left(N\left(m+n\right)\right)$. Finally, by plugging Eq. \eqref{eq:Fast Formula}
into Eq. \eqref{eq:Obj}, the parameter $\boldsymbol{\phi}$\textcolor{green}{{}
}can be numerically optimized by minimizing

\begin{equation}
\hat{\boldsymbol{\phi}}=\mathrm{arg}\underset{\boldsymbol{\phi}}{\min}\left\{ N\log\hat{\sigma}^{2}+\log\left|\boldsymbol{S}\right|\right\} .\label{eq:Kron-Obj}
\end{equation}
The use of the Kronecker product \citep{stegleEfficientInferenceMatrixvariate2011a}
significantly reduces the computational complexity of matrix inversion
and determinant computations, which is demonstrated on a dataset comprising
millions of outputs in the real case study discussed in Sec. \ref{sec:Real-Case-Study}.\textcolor{red}{{}
}

\subsection{Functional-Output Effect Decomposition \label{subsec:FO Effect Decom}
}

FOED provides theoretical support for variance decomposition in FOFANOVA
and reveals how input variables influence functional outputs, facilitating
the understanding of input-output relationships. Moreover, its additive
orthogonal structure improves both model interpretability and predictive
performance \citep{duvenaudAdditiveGaussianProcesses2011,durrandeAdditiveCovarianceKernels2012,sungMultiresolutionFunctionalANOVA2020},
providing a comprehensive understanding of complex black-box computer
experiments. 

Given a new input point $\boldsymbol{x}$ at output position $t$,
the model prediction $f\left(\boldsymbol{x},t\right)$ is given by
 
\[
\hat{f}\left(\boldsymbol{x},t\right)=\left(\delta_{t}^{2}\boldsymbol{k}_{t}\odot\left(\bigodot_{i=1}^{d}\left(\delta_{i}^{2}\boldsymbol{k}_{i}+\mathbf{1}_{N}\right)\right)\right)^{\top}\boldsymbol{\gamma},
\]
where $\boldsymbol{k}_{i}=\tilde{k}_{i}\left(x_{i},\boldsymbol{X}_{i}\right)=\left[\tilde{k}_{i}\left(x_{i},x_{ui}\right)\right]_{u\in\left\{ 1,\cdots,N\right\} }$,
$\boldsymbol{k}_{t}=k_{t}\left(t,\boldsymbol{T}\right)=\left[k_{t}\left(t,t_{v}\right)\right]_{v\in\left\{ 1,\cdots,N\right\} }$,
and $\boldsymbol{\gamma}=\boldsymbol{K}^{-1}\boldsymbol{y}$. The
model prediction utilizes the same Hadamard product structure of the
covariance matrix $\boldsymbol{K}$ in Eq. \eqref{eq:Cov Matrix},
and the time complexity of the model prediction is $O\left(Nd\right)$,
making the FOAGP prediction also comparable to the standard GP prediction.
The prediction $\hat{f}\left(\boldsymbol{x},t\right)$ can be easily
decomposed into $2^{d}$ effect predictions $\left\{ \hat{f}_{\boldsymbol{u}\mid t}\left(\boldsymbol{x}_{\boldsymbol{u}},t\right)\right\} $,
which is given by 
\begin{equation}
\hat{f}_{\boldsymbol{u}\mid t}\left(\boldsymbol{x}_{\boldsymbol{u}},t\right)=\left(\delta_{t}^{2}\boldsymbol{k}_{t}\odot\left(\bigodot_{i\in\boldsymbol{u}}\delta_{i}^{2}\boldsymbol{k}_{i}\right)\right)^{\top}\boldsymbol{\gamma}.\label{eq:Effect Pred}
\end{equation}
The time complexity of computing $\hat{f}_{\boldsymbol{u}\mid t}\left(\boldsymbol{x}_{\boldsymbol{u}},t\right)$
remains $O\left(Nd\right)$, which is identical to the time complexity
of model prediction.

When the functional-output outputs are collected at grid output position,
the prediction of $f\left(\boldsymbol{x},t\right)$ and $f_{\boldsymbol{u}\mid t}\left(\boldsymbol{x}_{\boldsymbol{u}},t\right)$
can be rewritten as 
\begin{equation}
\begin{cases}
\hat{f}\left(\boldsymbol{x},t\right) & =\left(\delta_{t}^{2}\boldsymbol{r}_{t}\otimes\left(\bigodot_{i=1}^{d}\left(\delta_{i}^{2}\boldsymbol{r}_{i}+\mathbf{1}_{m}\right)\right)\right)^{\top}\boldsymbol{\gamma}\\
\hat{f}_{\boldsymbol{u}\mid t}\left(\boldsymbol{x}_{\boldsymbol{u}},t\right) & =\left(\delta_{t}^{2}\boldsymbol{r}_{t}\otimes\left(\bigodot_{i\in\boldsymbol{u}}\delta_{i}^{2}\boldsymbol{r}_{i}\right)\right)^{\top}\boldsymbol{\gamma}
\end{cases},\label{eq:KronPred}
\end{equation}
where $\boldsymbol{r}_{t}=\left[k_{t}\left(t,\tau_{v}\right)\right]_{v\in\left\{ 1,\cdots,n\right\} }$
and $\boldsymbol{r}_{i}=\left[\tilde{k}_{i}\left(x_{i},\chi_{ui}\right)\right]_{u\in\left\{ 1,\cdots,m\right\} }$.
The time complexity of the computing $\hat{f}_{\boldsymbol{u}\mid t}\left(\boldsymbol{x}_{\boldsymbol{u}},t\right)$
is further reduced to $O\left(m\left(n+d\right)\right)$. 

Due to the absence of  input-output interactions, the effect predictions
$\left\{ \hat{f}_{\boldsymbol{u}\mid t}\left(\boldsymbol{x}_{\boldsymbol{u}},t\right)\right\} $
exactly result in orthogonal FOED, which is summarized in the following
Theorem \ref{thm:Effect Decomposition}. 
\begin{thm}
\label{thm:Effect Decomposition} The following properties of FOAGP
hold under the assumption that the elements in the vector $\left[\boldsymbol{x}^{\top},t\right]^{\top}$
are mutually independent: 

(a) Conditional zero mean: $\int\hat{f}_{\boldsymbol{u}\mid t}\left(\boldsymbol{x}_{\boldsymbol{u}},t\right)\mathrm{d}F\left(\boldsymbol{x}\right)=0,\forall\boldsymbol{u}\ne\varnothing,\forall t.$ 

(b) Conditional orthogonality: $\int\hat{f}_{\boldsymbol{u}\mid t}\left(\boldsymbol{x}_{\boldsymbol{u}},t\right)\hat{f}_{\boldsymbol{v}\mid t}\left(\boldsymbol{x}_{\boldsymbol{v}},t\right)\mathrm{d}F\left(\boldsymbol{x}\right)=0,\forall\boldsymbol{u}\ne\boldsymbol{v},\forall t,$
where $F\left(\boldsymbol{x}\right)$ is the distribution function
of $\boldsymbol{x}$. 
\end{thm}
The conditional zero mean of Theorem \ref{thm:Effect Decomposition}(a),
derived from the conditional orthogonal constraints in Eq. \eqref{eq:C-OCons},
ensures that the estimated effects $\left\{ \hat{f}_{\boldsymbol{u}\mid t}\left(\boldsymbol{x}_{\boldsymbol{u}},t\right)\mid\boldsymbol{u}\ne\varnothing\right\} $
have zero mean at any output position $t$, further enabling the estimated
mean effect $\hat{f}_{0}\left(t\right)$ accurately represents the
mean of the functional output $y\left(\boldsymbol{x},t\right)$. Additionally,
the conditional orthogonality of Theorem \ref{thm:Effect Decomposition}(b),
which directly follows from the conditional zero mean of Theorem \ref{thm:Effect Decomposition}(a)
and the independence assumption, ensures the identifiability of the
proposed FOAGP model, improves model interpretability, and facilitates
the following variance decomposition in Sec. \ref{subsec:Variance Decomposition}.
Furthermore, theorem \ref{thm:Effect Decomposition} demonstrates
that by embedding orthogonal FOED directly into the model structure,
FOAGP simultaneously performs FOED during modeling, thereby reducing
the time cost associated with the separate decomposition step. 

\subsection{\textcolor{black}{Global Local Variance Decomposition \label{subsec:Variance Decomposition} }}

\begin{figure}[h]
\begin{centering}
\includegraphics[width=9cm]{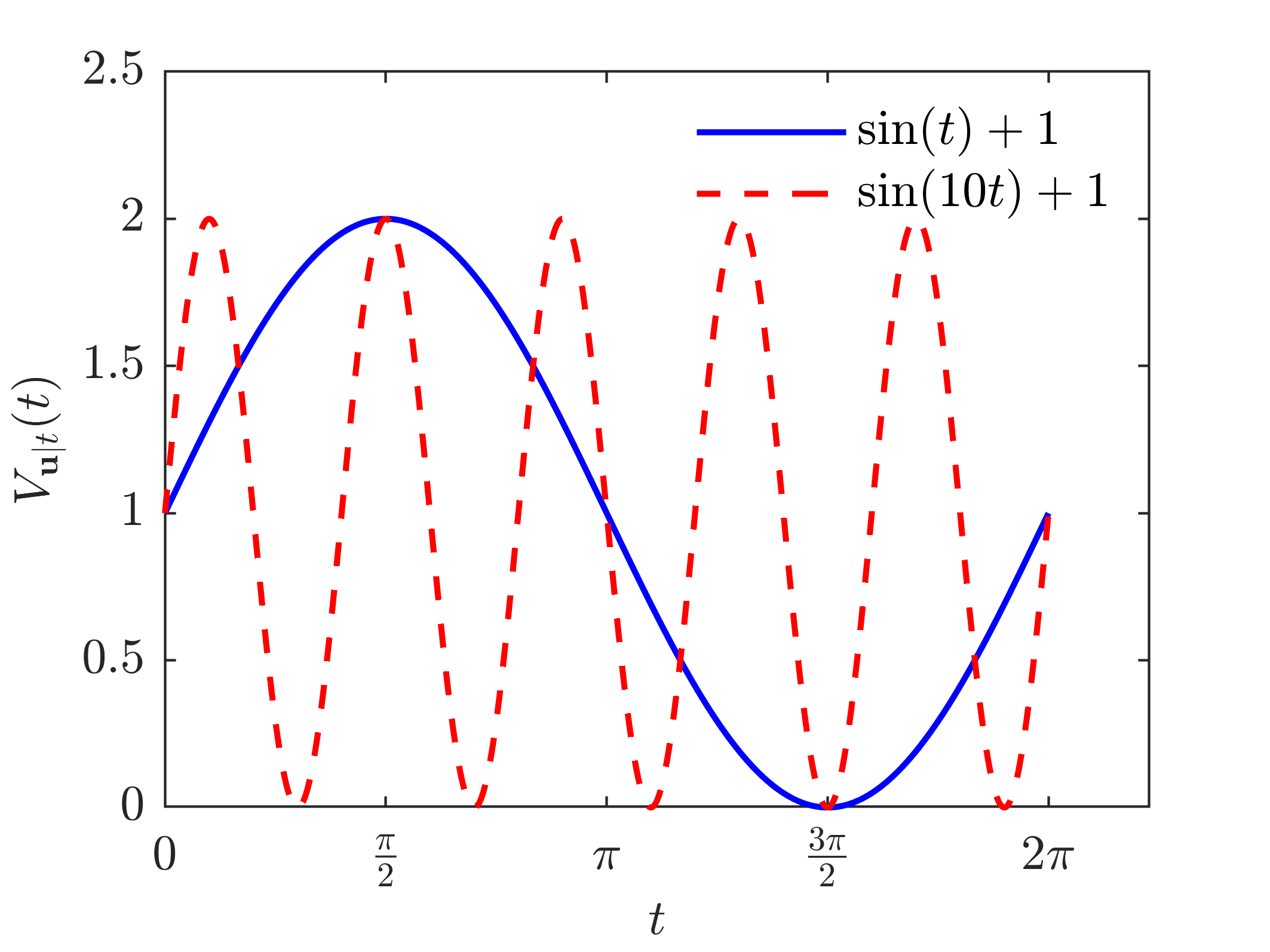}
\par\end{centering}
\caption{Two different local variance with the same global variance.\textcolor{red}{{}
\label{fig:Dynamic_vs_Global}}}
\end{figure}

Variance decomposition lies at the core of FOFANOVA, quantifying the
contribution of each input variable to the overall variability of
functional outputs. Local variance decomposition assesses input importance
at individual output positions, while global variance decomposition
evaluates contributions across the entire output domain. Leveraging
the full potential of orthogonal FOED in Theorem \ref{thm:Effect Decomposition},
the local variance decomposition and global variance decomposition
of FOAGP are established in the following Theorem \ref{thm:Variance Decomposition},
offering both local and global insights within a unified sensitivity
analysis framework. 
\begin{thm}
\label{thm:Variance Decomposition} The following properties of FOAGP
hold under the assumption that the elements in the vector $\left[\boldsymbol{x}^{\top},t\right]^{\top}$
are mutually independent: 

(a) Local variance decomposition: $\mathbb{V}_{\boldsymbol{x}}\left(\hat{f}\left(\boldsymbol{x},t\right)\right)=\sum_{\boldsymbol{u}\subseteq\mathcal{D}}\mathbb{V}_{\boldsymbol{x}_{\boldsymbol{u}}}\left(\hat{f}_{\boldsymbol{u}\mid t}\left(\boldsymbol{x}_{\boldsymbol{u}},t\right)\right),\forall t.$ 

(b) Global variance decomposition: $\mathbb{E}_{t}\left[\mathbb{V}_{\boldsymbol{x}}\left(\hat{f}\left(\boldsymbol{x},t\right)\right)\right]=\sum_{\boldsymbol{u}\subseteq\mathcal{D}}\mathbb{E}_{t}\left[\mathbb{V}_{\boldsymbol{x}_{\boldsymbol{u}}}\left(\hat{f}_{\boldsymbol{u}\mid t}\left(\boldsymbol{x}_{\boldsymbol{u}},t\right)\right)\right].$ 
\end{thm}
The local variance decomposition is validated through the conditional
mean and the conditional orthogonality in Theorem \ref{thm:Effect Decomposition}.
The global variance decomposition is then obtained by taking the expectation
with respect to output position $t$ of the local variance decomposition.
While the global variance decomposition provides an overall measure
of effect significance, the local variance decomposition provides
a finer-grained perspective. Consider the following example: Suppose
we have two local variance $\sin t+1$ and $\sin\left(10t\right)+1$
of two different effects over the interval $\left[0,2\pi\right]$,
as illustrated in Fig. \ref{fig:Dynamic_vs_Global}. The global variance
of both yields $1$, offering no distinction information between the
two different effects. This highlights that effects with different
local variances can share the same global variance. Integrating \ref{thm:Variance Decomposition}(a)
and Theorem \ref{thm:Variance Decomposition}(b) introduces a novel,
comprehensive approach to the variance decomposition of functional
outputs, which we refer to as global local variance decomposition.
Building on the global local variance decomposition, we then develop
the global local sensitivity analysis and deduce the analytical formulas
of the global local sensitivity indices along with their corresponding
estimators for practice. 

The local variances of $\hat{f}_{\boldsymbol{u}\mid t}\left(\boldsymbol{x}_{\boldsymbol{u}},t\right)$
and $\hat{f}\left(\boldsymbol{x},t\right)$ at output position $t$
are denoted by $V_{\boldsymbol{u}\mid t}\left(t\right)=\mathbb{V}_{\boldsymbol{x}_{\boldsymbol{u}}}\left(\hat{f}_{\boldsymbol{u}\mid t}\left(\boldsymbol{x}_{\boldsymbol{u}},t\right)\right)$
and $V_{t}\left(t\right)=\mathbb{V}_{\boldsymbol{x}}\left(\hat{f}\left(\boldsymbol{x},t\right)\right)$,
respectively.\textcolor{orange}{{} }The local Sobol' indices of $\boldsymbol{x}_{\boldsymbol{u}}$
at output position $t$ are defined as 
\begin{equation}
\mathcal{S}_{\boldsymbol{u}\mid t}\left(t\right)=\frac{V_{\boldsymbol{u}\mid t}\left(t\right)}{V_{t}\left(t\right)},\label{eq:Sobol}
\end{equation}
which are also called the local sensitivity indices. Due to the local
variance decomposition in Theorem \ref{thm:Variance Decomposition}(a),
we can express the denominator $V_{t}\left(t\right)$ in Eq. \eqref{eq:Sobol}
as $\sum_{\boldsymbol{u}\subseteq\mathcal{D}}V_{\boldsymbol{u}\mid t}\left(t\right)$.
Thus, the local Sobol' indices $\mathcal{S}_{\boldsymbol{u}\mid t}\left(t\right)$
falls within the range $\left[0,1\right]$. Additionally, to obtain
the local Sobol' indices $\mathcal{S}_{\boldsymbol{u}\mid t}\left(t\right)$,
we only need to compute $V_{\boldsymbol{u}\mid t}\left(t\right)$
and normalize them so that their sum equals 1.  

The ECV of $\hat{f}_{\boldsymbol{u}\mid t}\left(\boldsymbol{x}_{\boldsymbol{u}},t\right)$
and $\hat{f}\left(\boldsymbol{x},t\right)$ are defined as $V_{\boldsymbol{u}}=\mathbb{E}_{t}\left[V_{\boldsymbol{u}\mid t}\left(t\right)\right]$
and $V=\mathbb{E}_{t}\left[V_{t}\left(t\right)\right]$, respectively.
The ECV averages the local variance over the whole output domain,
so we also call it the global variance. The ECV sensitivity indices
\citep{drigneiFunctionalANOVAComputer2010} of $\boldsymbol{x}_{\boldsymbol{u}}$
are defined as 
\begin{equation}
\mathcal{S}_{\boldsymbol{u}}=\frac{V_{\boldsymbol{u}}}{V},\label{eq:Global Indices}
\end{equation}
which we also refer to as the global sensitivity indices in this study.
Similar to the local sensitivity analysis, by leveraging the global
variance decomposition in Theorem \ref{thm:Variance Decomposition}(b),
we can express the denominator $V$ in Eq. \eqref{eq:Global Indices}
as $\sum_{\boldsymbol{u}\subseteq\mathcal{D}}V_{\boldsymbol{u}}$.
Again, we only need to compute the ECV $V_{\boldsymbol{u}}$ of each
estimated effect $\hat{f}_{\boldsymbol{u}\mid t}\left(\boldsymbol{x}_{\boldsymbol{u}},t\right)$,
and then normalize them so that their sum equals 1, completing the
calculation of the ECV sensitivity indices $\mathcal{S}_{\boldsymbol{u}}$. 

Within the FOAGP framework, the conditional variance $V_{\boldsymbol{u}\mid t}\left(t\right)$
and the global variance $V_{\boldsymbol{u}}$ can be derived analytically,
which is summarized in the following theorem \ref{thm:Analytic}. 
\begin{thm}
\label{thm:Analytic} Under the assumption that the elements in the
vector $\left[\boldsymbol{x}^{\top},t\right]^{\top}$ are mutually
independent, the local variance $V_{\boldsymbol{u}\mid t}\left(t\right)$
in Eq. \eqref{eq:Sobol} and the global variance $V_{\boldsymbol{u}}$
in Eq. \eqref{eq:Global Indices} have analytical formulas within
the FOAGP framework, that is, 
\begin{equation}
V_{\boldsymbol{u}\mid t}\left(t\right)=\boldsymbol{\gamma}^{\top}\left(\left(\delta_{t}^{4}\boldsymbol{k}_{t}\boldsymbol{k}_{t}^{\top}\right)\odot\left(\bigodot_{i\in\boldsymbol{u}}\delta_{i}^{4}\mathbb{E}_{x_{i}}\left[\boldsymbol{k}_{i}\boldsymbol{k}_{i}^{\top}\right]\right)\right)\boldsymbol{\gamma},\label{eq:Analytic Dynamic Var}
\end{equation}
and 
\begin{equation}
V_{\boldsymbol{u}}=\boldsymbol{\gamma}^{\top}\left(\delta_{t}^{4}\mathbb{E}_{t}\left[\boldsymbol{k}_{t}\boldsymbol{k}_{t}^{\top}\right]\odot\left(\bigodot_{i\in\boldsymbol{u}}\delta_{i}^{4}\mathbb{E}_{x_{i}}\left[\boldsymbol{k}_{i}\boldsymbol{k}_{i}^{\top}\right]\right)\right)\boldsymbol{\gamma},\label{eq:Analytic Global Var}
\end{equation}
where $\mathbb{E}_{t}\left[\boldsymbol{k}_{t}\boldsymbol{k}_{t}^{\top}\right]=\int\boldsymbol{k}_{t}\boldsymbol{k}_{t}^{\top}\mathrm{d}F_{t}\left(t\right)$
and $F_{t}\left(t\right)$ is the CDF of $t$. 
\end{thm}
Noting that $\mathbb{E}_{x_{i}}\left[\boldsymbol{k}_{i}\boldsymbol{k}_{i}^{\top}\right]$
and $\mathbb{E}_{t}\left[\boldsymbol{k}_{t}\boldsymbol{k}_{t}^{\top}\right]$
may not have closed forms due to the unknown distribution of the input
$\boldsymbol{x}$ and output position $t$, the conditional variance
$V_{\boldsymbol{u}\mid t}\left(t\right)$ and the global variance
$V_{\boldsymbol{u}}$ are difficult to compute directly. In practice
we can estimate these variance through the following two estimators
\begin{equation}
\begin{cases}
\hat{V}_{\boldsymbol{u}\mid t}\left(t\right) & =\boldsymbol{\gamma}^{\top}\left(\left(\delta_{t}^{4}\boldsymbol{k}_{t}\boldsymbol{k}_{t}^{\top}\right)\odot\left(\bigodot_{i\in\boldsymbol{u}}\left(\delta_{i}^{4}\boldsymbol{K}_{i}^{2}/N\right)\right)\right)\boldsymbol{\gamma}\\
\hat{V}_{\boldsymbol{u}} & =\boldsymbol{\gamma}^{\top}\left(\left(\delta_{t}^{4}\boldsymbol{K}_{t}^{2}/N\right)\odot\left(\bigodot_{i\in\boldsymbol{u}}\left(\delta_{i}^{4}\boldsymbol{K}_{i}^{2}/N\right)\right)\right)\boldsymbol{\gamma}
\end{cases},\label{eq:VarEstimator}
\end{equation}
which uses the training dataset as an empirical distribution. This
approach allows for computing sensitivity indices just after model
fitting, making FOAGP a consistent and handy sensitivity analysis
tool for functional-output computer experiments. Additionally, all
estimators for sensitivity analysis rely solely on the training dataset.
These estimators utilize the training dataset to compute local sensitivity
indices and global sensitivity indices, enabling global local sensitivity
analysis without the prior knowledge about the underlying data distribution. 

When the functional outputs are collected at grid output position,
the corresponding analytical formulas and estimators can be also derived
as follows: 
\begin{equation}
\begin{cases}
V_{\boldsymbol{u}\mid t}\left(t\right) & =\mathrm{tr}\left(\boldsymbol{\varGamma}^{\top}\left(\delta_{t}^{4}\boldsymbol{r}_{t}\boldsymbol{r}_{t}^{\top}\right)\boldsymbol{\varGamma}\left(\bigodot_{i\in\boldsymbol{u}}\delta_{i}^{4}\mathbb{E}_{x_{i}}\left[\boldsymbol{\boldsymbol{r}}_{i}\boldsymbol{r}_{i}^{\top}\right]\right)\right)\\
\hat{V}_{\boldsymbol{u}\mid t}\left(t\right) & =\mathrm{tr}\left(\boldsymbol{\varGamma}^{\top}\left(\delta_{t}^{4}\boldsymbol{r}_{t}\boldsymbol{\boldsymbol{r}}_{t}^{\top}\right)\boldsymbol{\varGamma}\left(\bigodot_{i\in\boldsymbol{u}}\left(\delta_{i}^{4}\boldsymbol{R}_{i}^{2}/m\right)\right)\right)\\
V_{\boldsymbol{u}} & =\mathrm{tr}\left(\boldsymbol{\varGamma}^{\top}\left(\delta_{t}^{4}\mathbb{E}_{t}\left[\boldsymbol{\boldsymbol{r}}_{t}\boldsymbol{r}_{t}^{\top}\right]\right)\boldsymbol{\varGamma}\left(\bigodot_{i\in\boldsymbol{u}}\delta_{i}^{4}\mathbb{E}_{x_{i}}\left[\boldsymbol{r}_{i}\boldsymbol{\boldsymbol{r}}_{i}^{\top}\right]\right)\right)\\
\hat{V}_{\boldsymbol{u}} & =\mathrm{tr}\left(\boldsymbol{\varGamma}^{\top}\left(\delta_{t}^{4}\boldsymbol{R}_{t}^{2}/n\right)\boldsymbol{\varGamma}\left(\bigodot_{i\in\boldsymbol{u}}\left(\delta_{i}^{4}\boldsymbol{R}_{i}^{2}/m\right)\right)\right)
\end{cases},\label{eq:KronVarEstimator}
\end{equation}
where $\boldsymbol{\varGamma}$ is the matricization of $\boldsymbol{\gamma}$
\citep{zhangMatrixAnalysisApplications2017}. Given that the trace
operation depends only on the diagonal elements of a matrix, the full
result of the matrix multiplication is not necessary. The computational
efficiency can be significantly enhanced by focusing only on the diagonal
elements. 

Since the local Sobol' indices in Eq. \eqref{eq:Sobol} remain invariant
under the bijective transformation \citep{owenShapleyValueMeasuring2017},
the global sensitivity indices derived based on them also retain this
property. Therefore, the numerical stability of the FOAGP framework
can be further enhanced by utilizing any bijective transformation
of the data before training, such as utilizing the normalizing flow
\citep{papamakariosNormalizingFlowsProbabilistic2021} or mapping
the input spaces to $\left[0,1\right]^{d+1}$ with a uniform distribution.
The latter transformation will be applied to the real case study discussed
in Sec. \ref{sec:Real-Case-Study}. 

While ECV sensitivity indices derived from global variance decomposition
provide an overall measure of effect significance, local Sobol' indices
based on local variance decomposition provide detailed insight into
the local effect significance, enabling a more comprehensive sensitivity
analysis. By integrating these two types of sensitivity indices, we
refer to this approach as global local sensitivity analysis. Combined
with the global local sensitivity analysis, FOAGP forms a unified
data-driven implementation of FOFANOVA for complex black-box computer
experiments. 

\section{Simulation Study \label{sec:Numerical-Examples}}

\textcolor{black}{In this section, we apply} FOAGP\textcolor{black}{{}
on two simulations to demonstrate its efficiency in orthogonal FOED
and variance decomposition. }Specifically, we compare the FOED of
FOAGP wi\textcolor{black}{th HDMR in Eq. \eqref{eq:HDMR}, where Legendre
polynomials---orthogonal polynomials on $\left[-1,1\right]$---are
used as the basis functions.}\textcolor{red}{{} } Then, we use FOAGP
to perform variance decomposition and compare the results with theoretical
variance decomposition. The mean trend is extracted for numerical
optimization of FOAGP \citep{burtRatesConvergenceSparse2019,katzfussVecchiaApproximationsGaussianProcess2020},
which is also aligned with the set of HDMR \citep{chengTimevariantReliabilityAnalysis2019}.
For HDMR, the input spaces are mapped to $\left[-1,1\right]^{3}$
with a uniform distribution before training, using inverse of CDF,
scaling, and translation. 
\begin{example}
\noindent \textbf{\textit{}}  To understand how FOAGP works, we
first consider the toy example $f\left(x_{1},x_{2},t\right)=1+2t+x_{1}t+2x_{2}t+x_{1}x_{2}t$.
To facilitate theoretical calculations and enable comparison with
the fitted results, the input space is set to be $\mathbb{R}^{3}$
with a standard ternary normal distribution. White noise $\epsilon\sim\mathcal{N}\left(0,0.1^{2}\right)$
is added to the data generation model. Example 1 is linear in each
input variable and interaction, the importance of each effect directly
proportional to its coefficients. Given the input variables follow
the same distribution, the variable $x_{2}$ with a larger coefficient
should have a higher have a higher variance and consequently higher
sensitivity indices. 
\end{example}
\noindent The theoretical FOED is easy to derive because each term
involving $x$'s in Example 1 is an odd function with respect to each
input variable $x_{i}$: $f_{0}(t)=1+2t,f_{1}(x_{1},t)=x_{1}t,f_{2}(x_{2},t)=2x_{2}t,f_{12}(x_{1},x_{2},t)=x_{1}x_{2}t.$
In addition, the local variance and ECV sensitivity indices can be
easily computed: $\ensuremath{V_{1\mid t}(t)=t^{2},V_{2\mid t}(t)=4t^{2},V_{12\mid t}(t)=t^{2},\mathcal{S}_{1}=0.1667,\mathcal{S}_{2}=0.6667,\mathcal{S}_{12}=0.1667.}$
These results are consistent with the previous analysis. 
\begin{example}
\noindent \textbf{\textit{}}   To further validate the capability
of FOAGP, we duplicate the example $f\left(x_{1},x_{2},t\right)=\left(t+1\right)\mathrm{e}^{-x_{1}t}\sin\left(2\pi t/x_{2}\right)$
given by \citet{drigneiFunctionalANOVAComputer2010}. The input space
is $\left(x_{1},x_{2},t\right)\in\left[1.0,2.0\right]\times\left[0.9,1.1\right]\times\left[0.2,2.0\right]$
with a uniform distribution and white noise $\epsilon\sim\mathcal{N}\left(0,0.01^{2}\right)$
is added to the data generation model. Example 2 exhibits oscillations
with decreasing amplitude. The second term is an exponential decay
function, where $x_{1}$ and $t$ have a multiplicative effect on
the amplitude. The third term is a sinusoidal function of $t$ and
$x_{2}$, with the frequency determined by $x_{2}$. Since that the
function is rapidly decreasing with respect to $x_{1}$ and oscillatory
with respect to $x_{2}$, we can expect that the ECV sensitivity index
of $x_{2}$ will be greater than that of $x_{1}$. Since $x_{1}$
and $x_{2}$ are in separate product terms, their interaction is likely
to be minimal, implying that the ECV sensitivity index $\mathcal{S}_{12}$
of the interaction effect should be small. 
\end{example}
\noindent Due to the complex structure of Example 2 and non-symmetric
input distribution, the theoretical FOED is not trivial and indeed
very complicated, not to mention the local variance decomposition.
They are computed by Wolfram Mathematica directly and are not presented
here due to space limitations.  Based on these results, the theoretical
ECV sensitivity indices are obtained: $\mathcal{S}_{1}=0.3251,\mathcal{S}_{2}=0.6027,\mathcal{S}_{12}=0.0722.$
These results are consistent with the previous analysis. Directly
applying the traditional FANOVA, \citet{drigneiFunctionalANOVAComputer2010}
obtained $\hat{\phi}_{1}=0.0003,\hat{\phi}_{2}=0.0003,\hat{\phi}_{t}=0.8749,\hat{\phi}_{12}=0.00004,\hat{\phi}_{1t}=0.0389,\hat{\phi}_{2t}=0.0750,\hat{\phi}_{12t}=0.0105$.
Clearly the output position $t$ dominates the the variance, overshadowing
the importance of the true input variables $x_{1}$ and $x_{2}$.
Through the relationship between Sobol' indices and ECV sensitivity
indices in Eq. \eqref{eq:ECV Indices based on FANOVA}, \citet{drigneiFunctionalANOVAComputer2010}
estimated ECV sensitivity indices by $\hat{\mathcal{S}}_{1}=0.31,\hat{\mathcal{S}}_{2}=0.60,\hat{\mathcal{S}}_{12}=0.09$.

\subsection{Effect Decomposition }

\begin{figure*}[t]
\begin{centering}
\subfloat[\label{fig:toy-f0}]{\begin{centering}
\includegraphics[width=5.2cm]{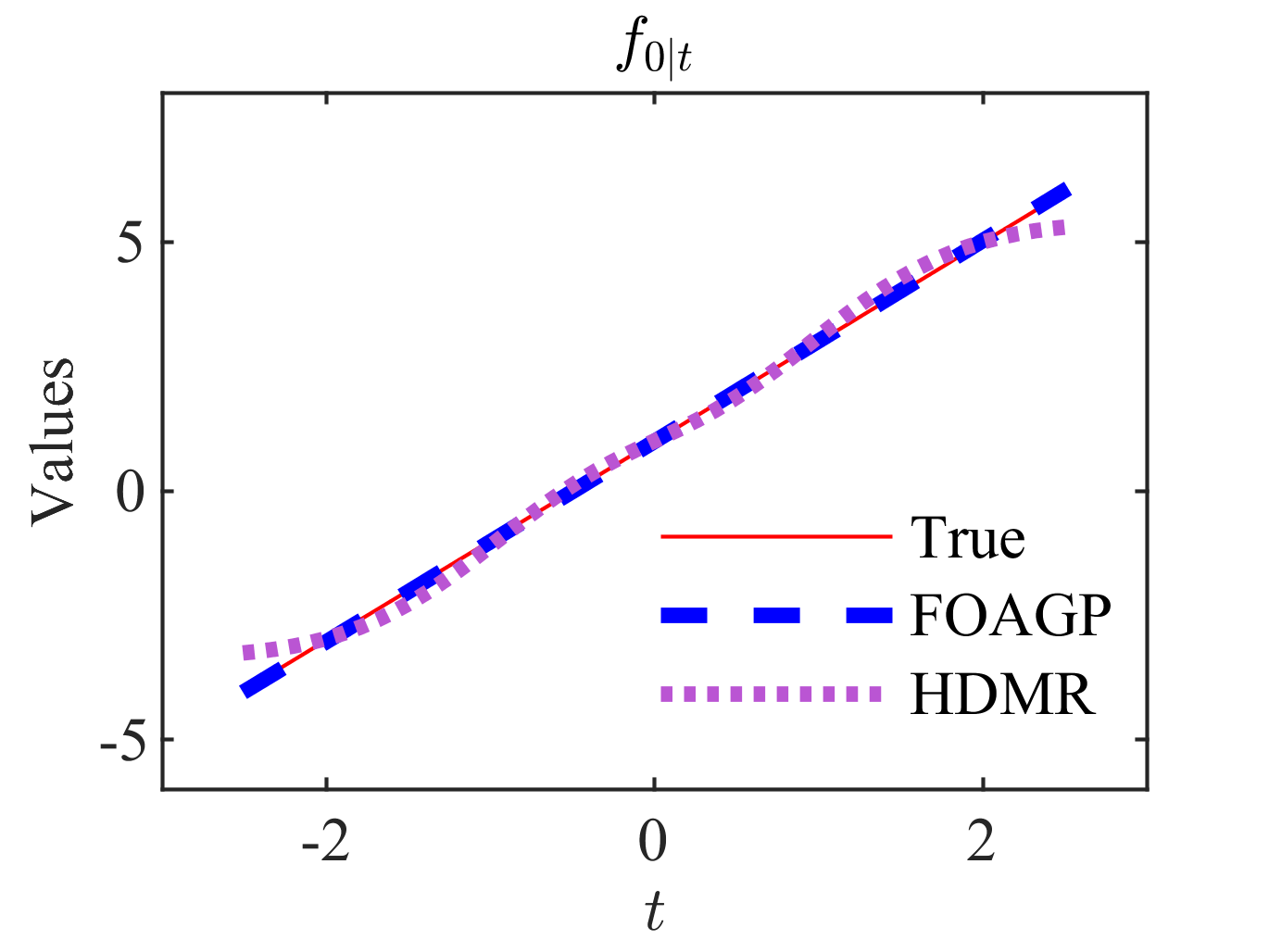}
\par\end{centering}
}\subfloat[\label{fig:toy-f1}]{\begin{centering}
\includegraphics[width=5.2cm]{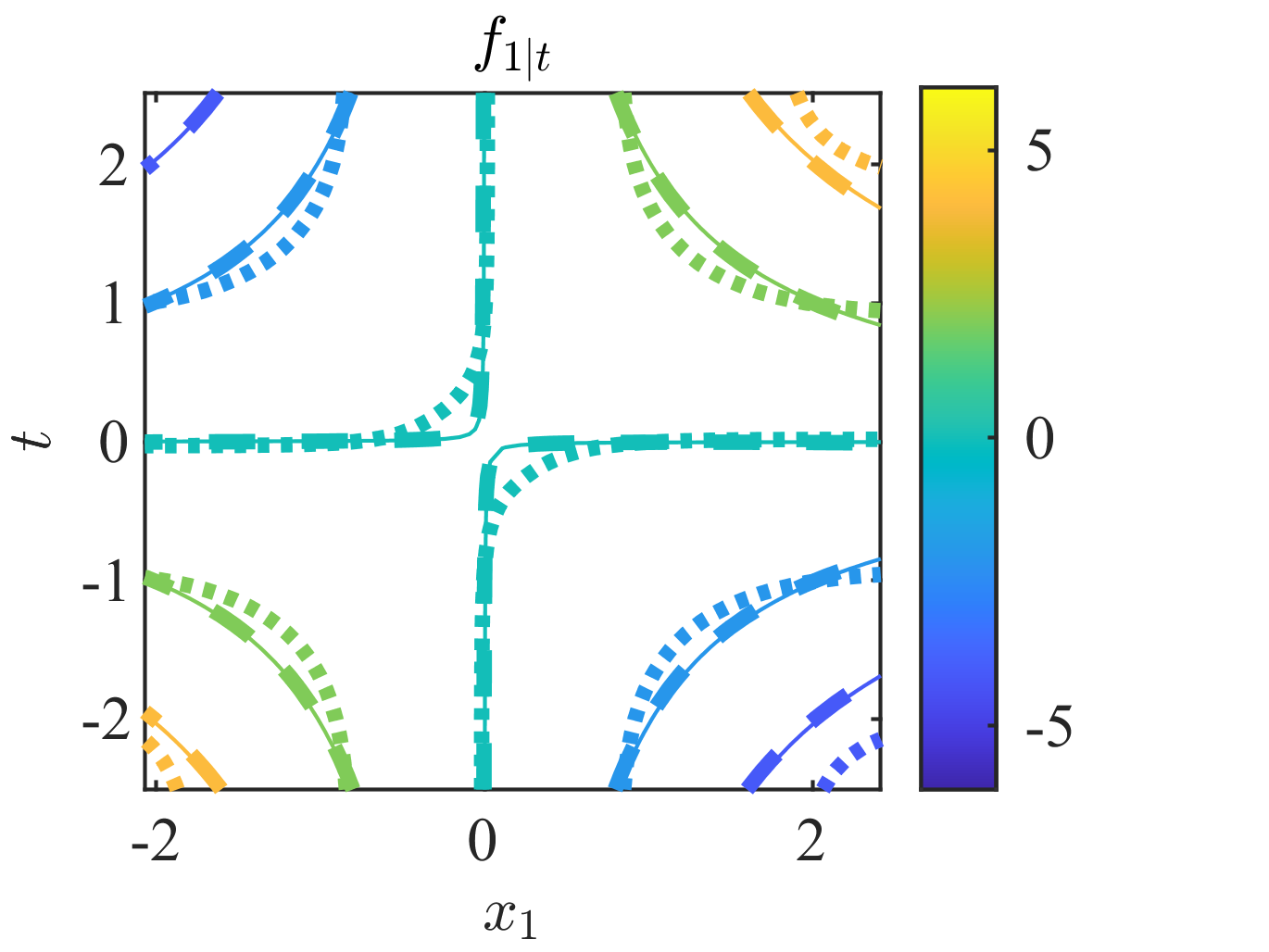}
\par\end{centering}
}\subfloat[\label{fig:toy-f2}]{\begin{centering}
\includegraphics[width=5.2cm]{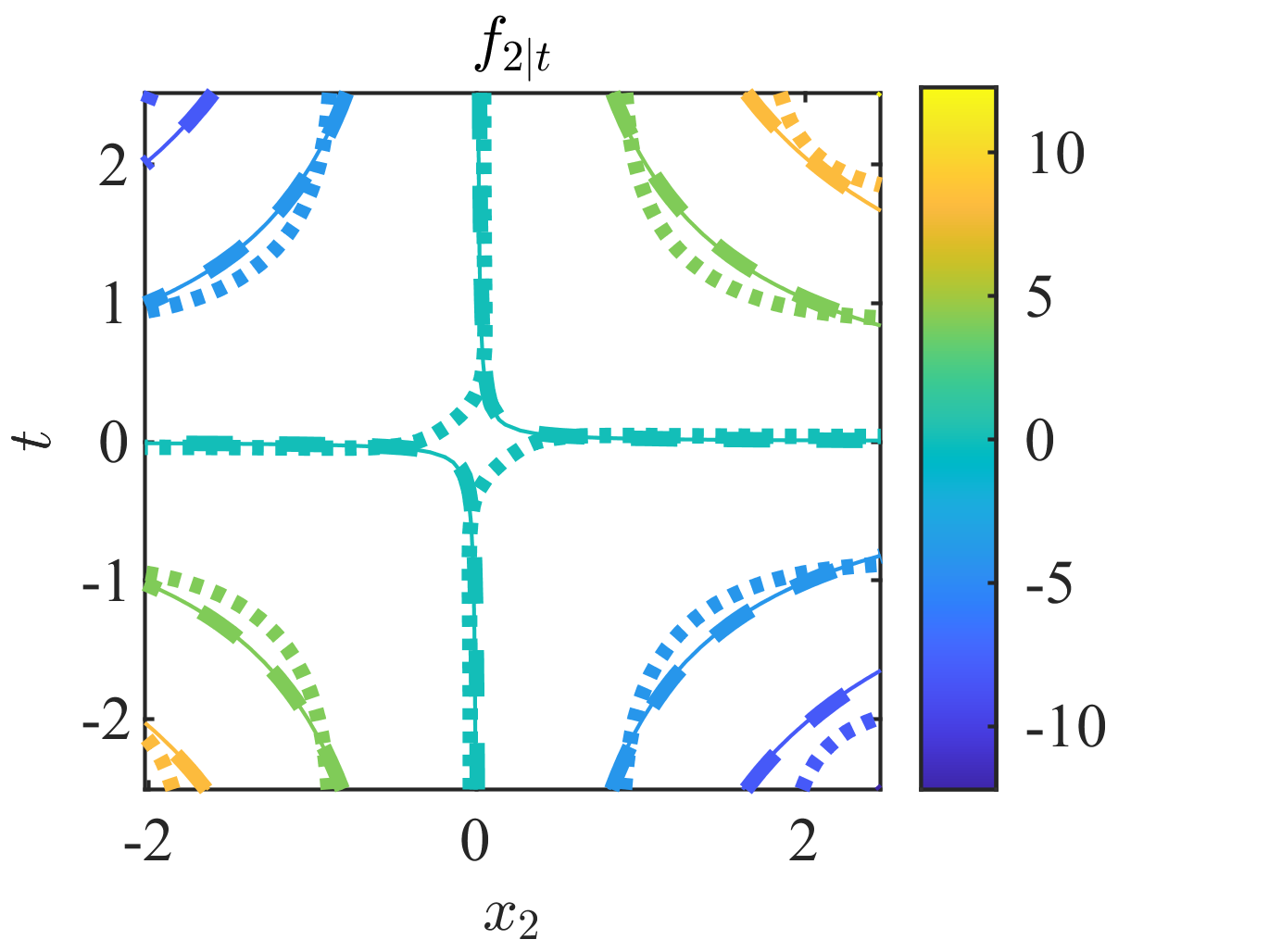}
\par\end{centering}
}
\par\end{centering}
\caption{Illustration of FOED of Example 1 using FOAGP and HDMR. The theoretical
values are depicted by thin solid lines; the predicted values of FOAGP
are depicted by dashed lines; the predicted values of HDMR are depicted
by dotted lines. \label{fig:toy-decomposition}}
\end{figure*}

\begin{figure*}[t]
\begin{centering}
\subfloat[\label{fig:2010-f0}]{\begin{centering}
\includegraphics[width=5.2cm]{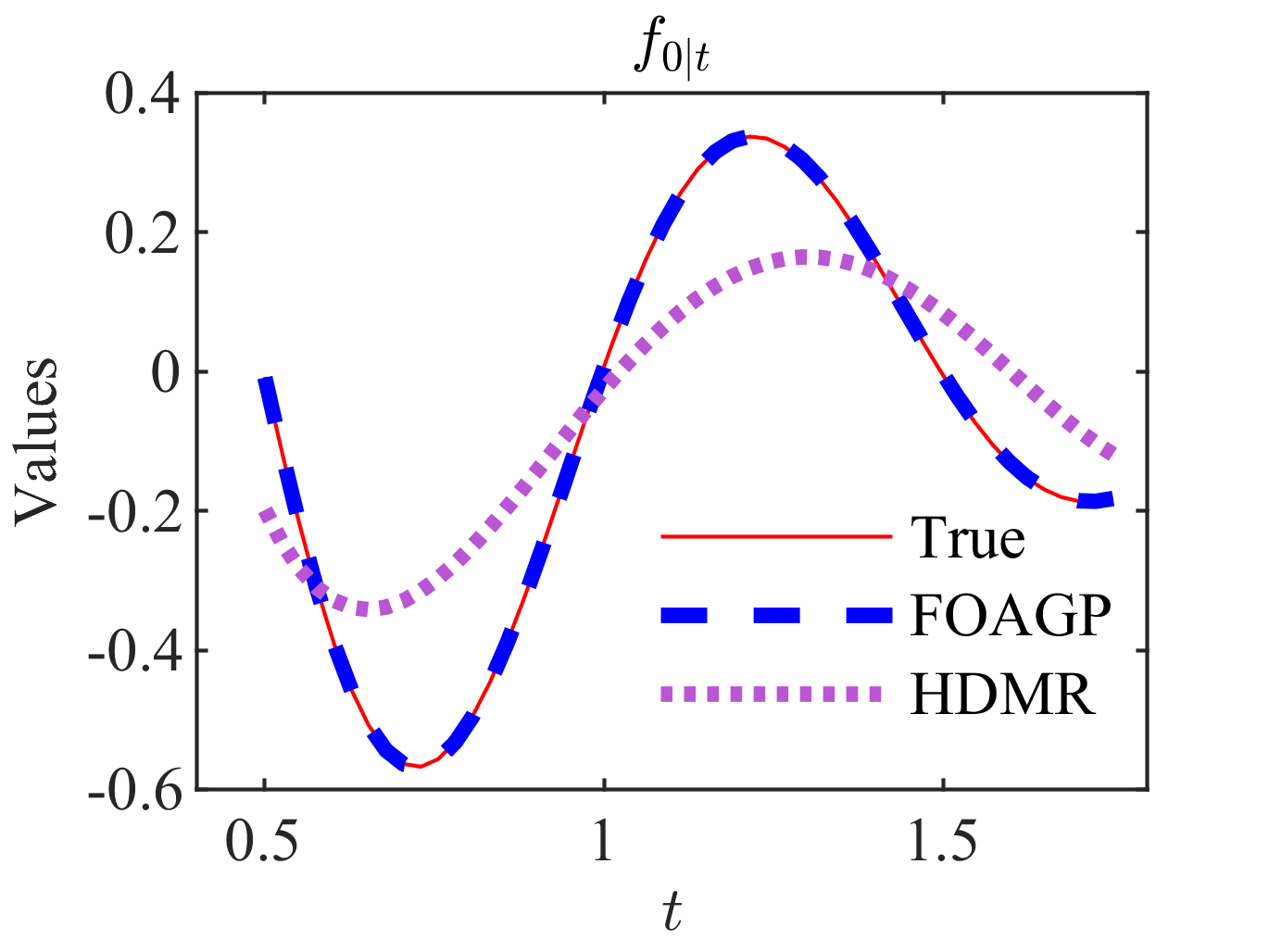}
\par\end{centering}
}\subfloat[\label{fig:2010-f1}]{\begin{centering}
\includegraphics[width=5.2cm]{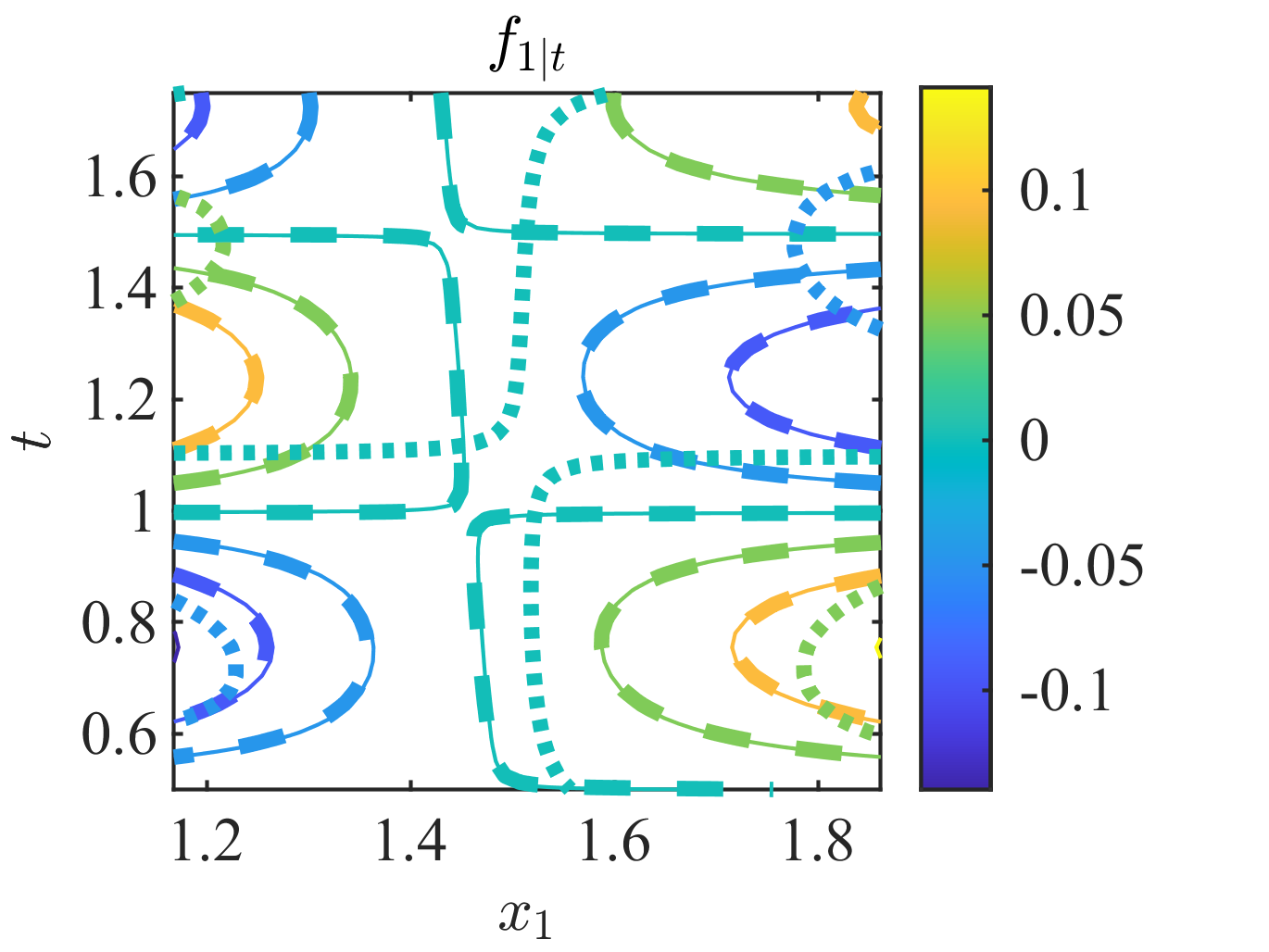}
\par\end{centering}
}\subfloat[\label{fig:2010-f2}]{\begin{centering}
\includegraphics[width=5.2cm]{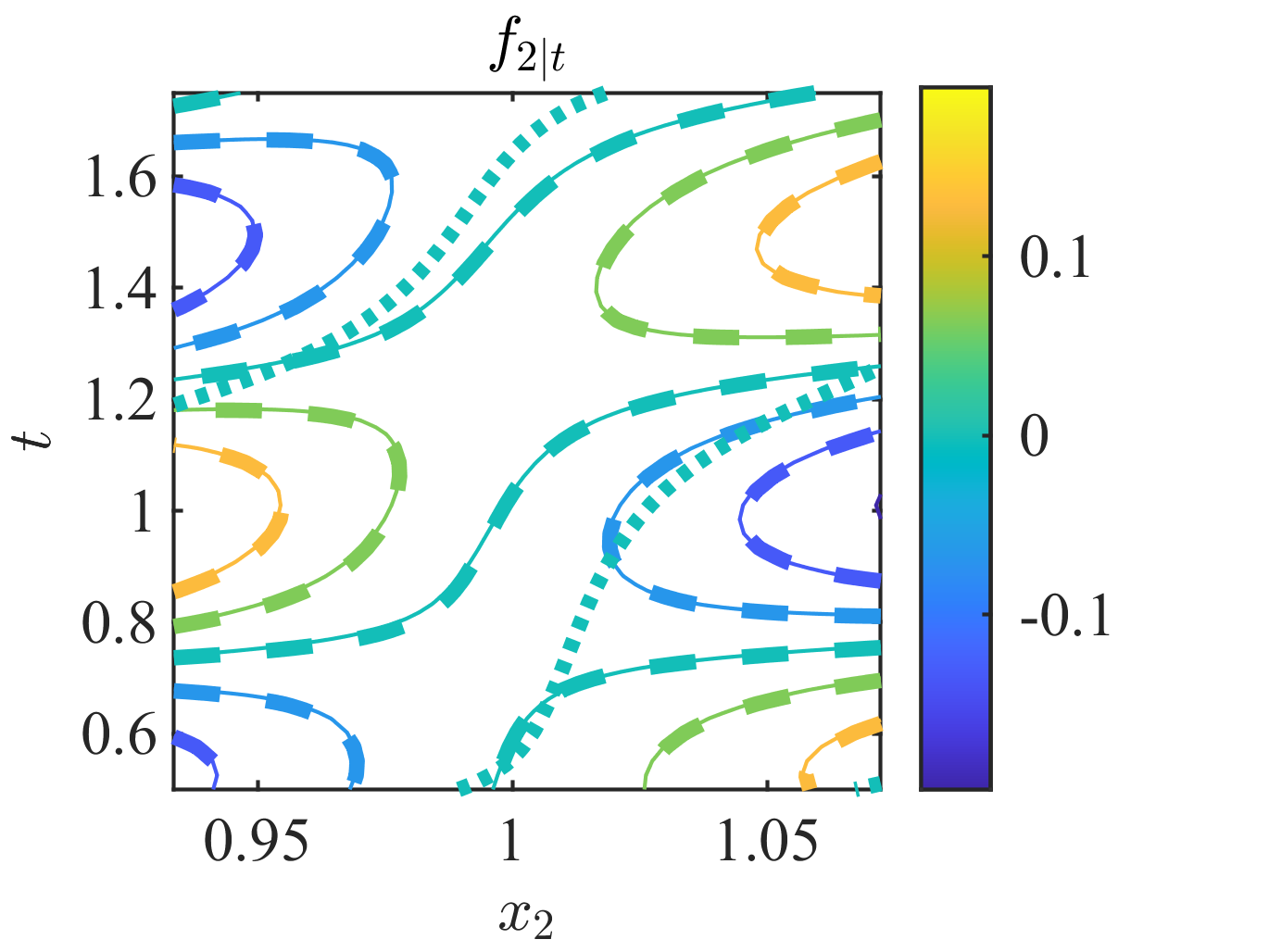}
\par\end{centering}
}
\par\end{centering}
\caption{Illustration of FOED of Example 2 using FOAGP and HDMR. The theoretical
values are depicted by thin solid lines; the predicted values of FOAGP
are depicted by dashed lines; the predicted values of HDMR are depicted
by dotted lines.\label{fig:2010-deccomposition}}
\end{figure*}

The FOED results using the FOAGP and HDMR are illustrated in the Fig.
\ref{fig:toy-decomposition} and Fig. \ref{fig:2010-deccomposition}.
We present the main effects, omitting the interaction effect due to
the limitations of the three-dimensional plot. The FOED of FOAGP closely
matches the theoretical FOED in both examples, providing a detailed
breakdown of functional outputs, while HDMR struggles to fit the theoretical
FOED. 

The primary reason for the relatively poor performance of HDMR lies
in its reliance on predefined basis functions with the truncated order
and its parametric structure, which limit its ability to model complex
nonlinear relationships. In Example 1, although the function appears
simple, the distribution transformation involving the inverse CDF
of $\mathcal{N}\left(0,1\right)$ introduces additional complexity
to the function, making it difficult for HDMR to accurately capture
the intricate relationships. In contrast, the proposed nonparametric
FOAGP model, which does not rely on prior knowledge about the underlying
data distribution, is more flexible and able to of effectively model
complex nonlinear relationships. 
 
\subsection{Local Variance Decomposition\textcolor{black}{{} }}

\begin{figure*}[t]
\noindent \begin{centering}
\subfloat[$\hat{V}_{1\mid t}$ and $V_{1\mid t}$\label{fig:toy-v1}]{\begin{centering}
\includegraphics[width=5.2cm]{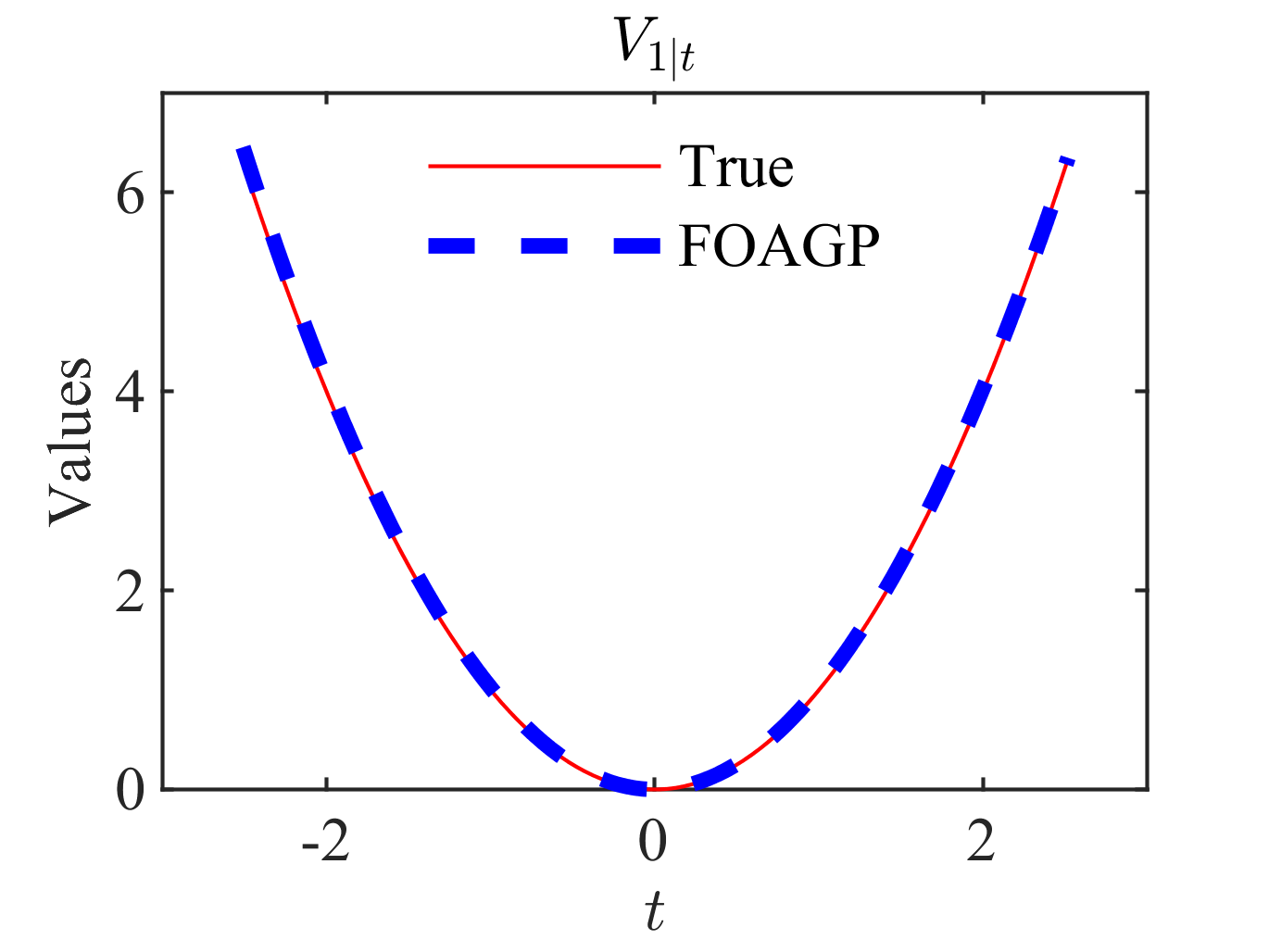}
\par\end{centering}
}\subfloat[$\hat{V}_{2\mid t}$ and $V_{2\mid t}$\label{fig:toy-v2}]{\begin{centering}
\includegraphics[width=5.2cm]{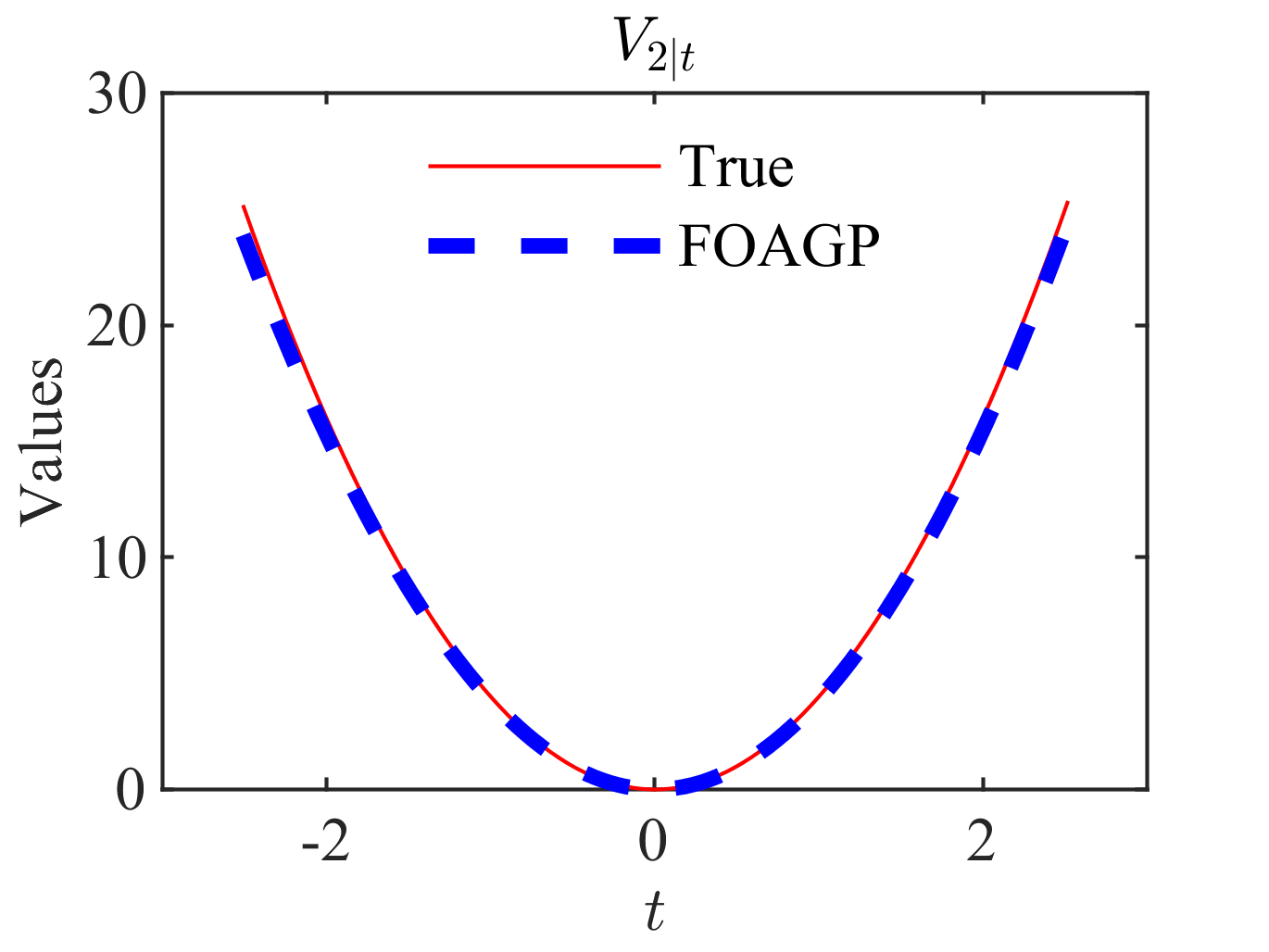}
\par\end{centering}
}\subfloat[$\hat{V}_{12\mid t}$ and $V_{12\mid t}$\label{fig:toy-v12}]{\begin{centering}
\includegraphics[width=5.2cm]{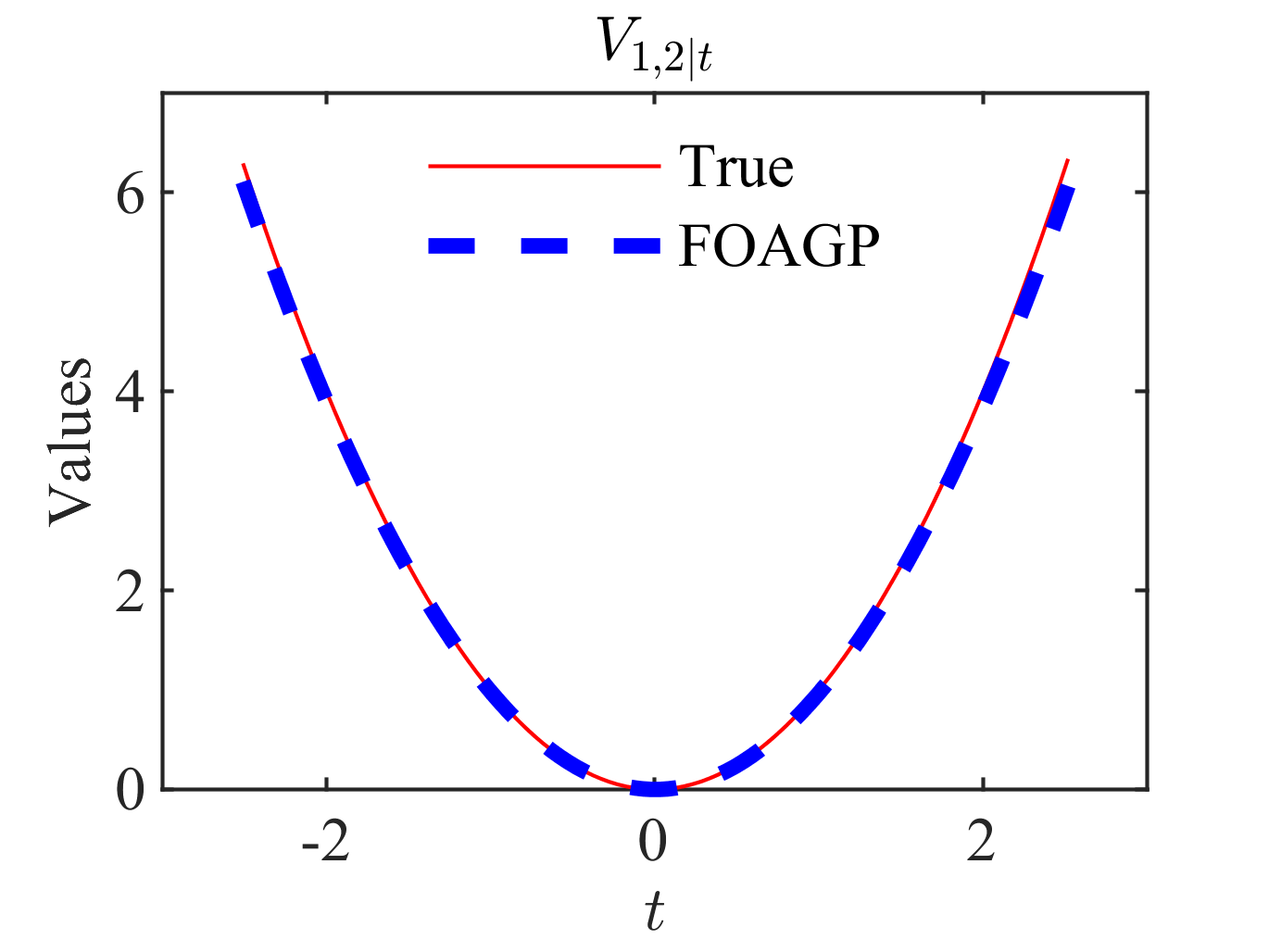}
\par\end{centering}
}
\par\end{centering}
\caption{Illustration of the local variance decomposition of Example 1 using
FOAGP. \label{fig:toy-variance}}
\end{figure*}

\begin{figure*}[t]
\noindent \begin{centering}
\subfloat[$\hat{V}_{1\mid t}$ and $V_{1\mid t}$\label{fig:2010-v1}]{\begin{centering}
\includegraphics[width=5.2cm]{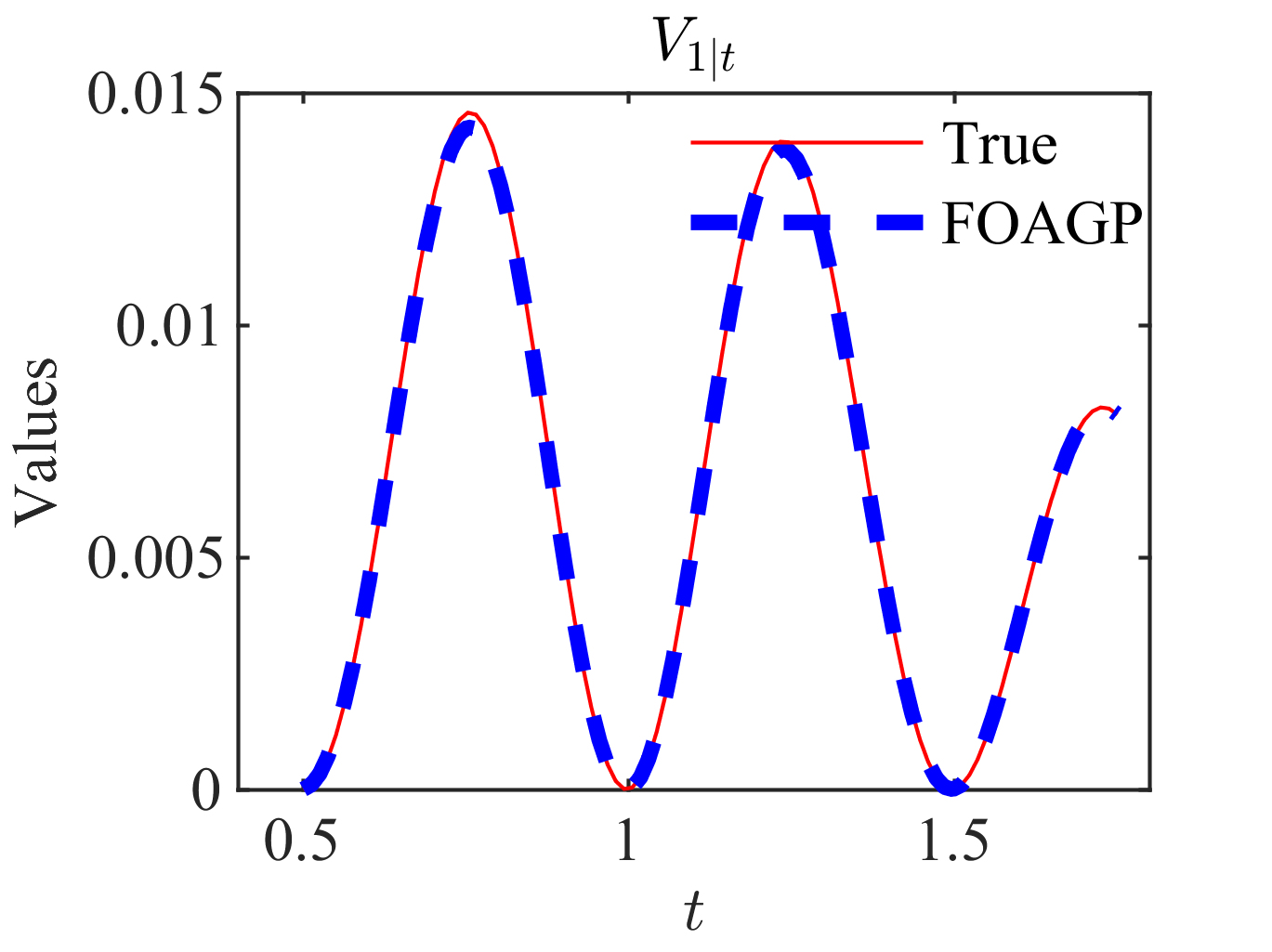}
\par\end{centering}
}\subfloat[$\hat{V}_{2\mid t}$ and $V_{2\mid t}$\label{fig:2010-v2}]{\begin{centering}
\includegraphics[width=5.2cm]{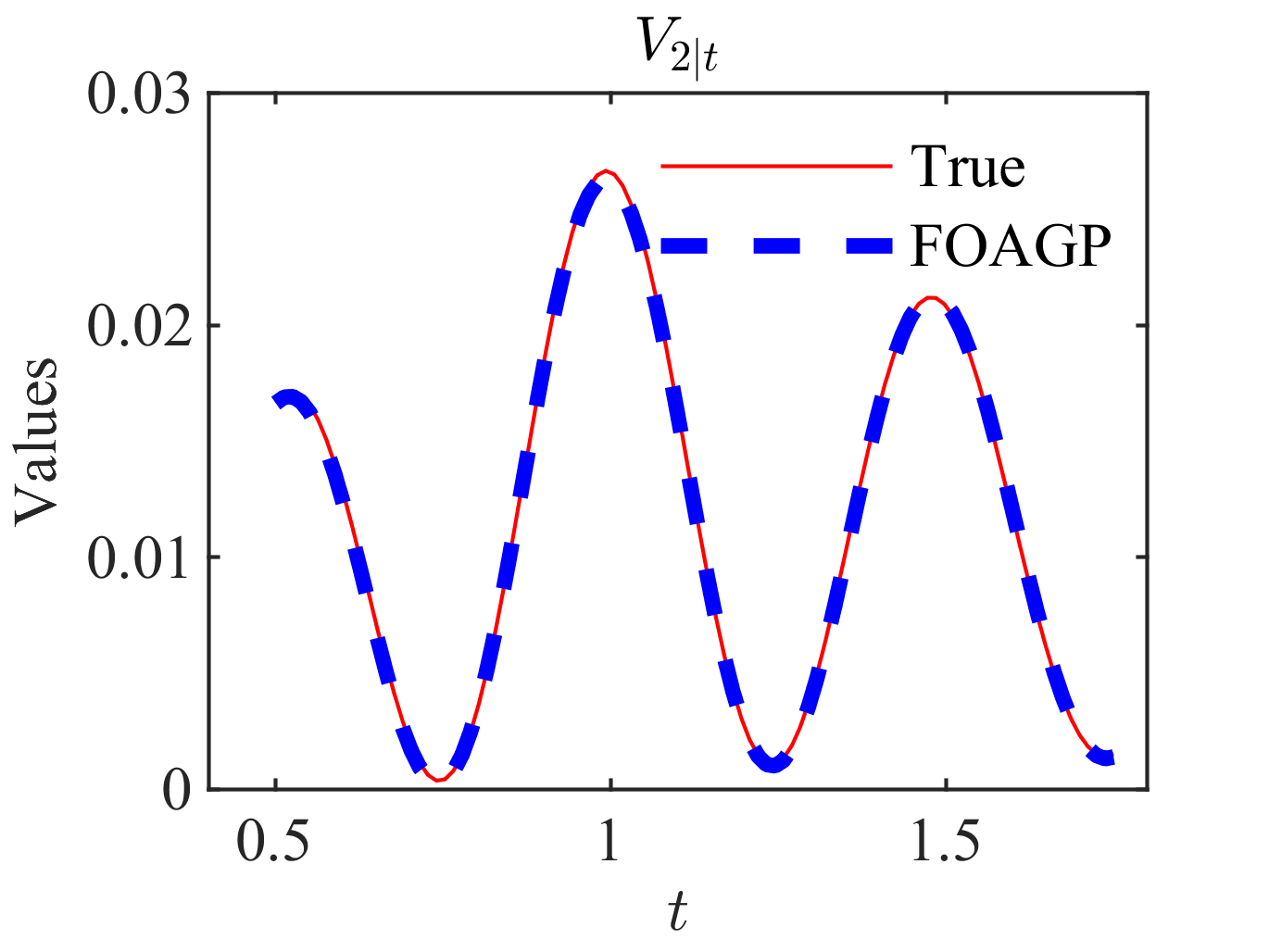}
\par\end{centering}
}\subfloat[$\hat{V}_{12\mid t}$ and $V_{12\mid t}$\label{fig:2010-v12}]{\begin{centering}
\includegraphics[width=5.2cm]{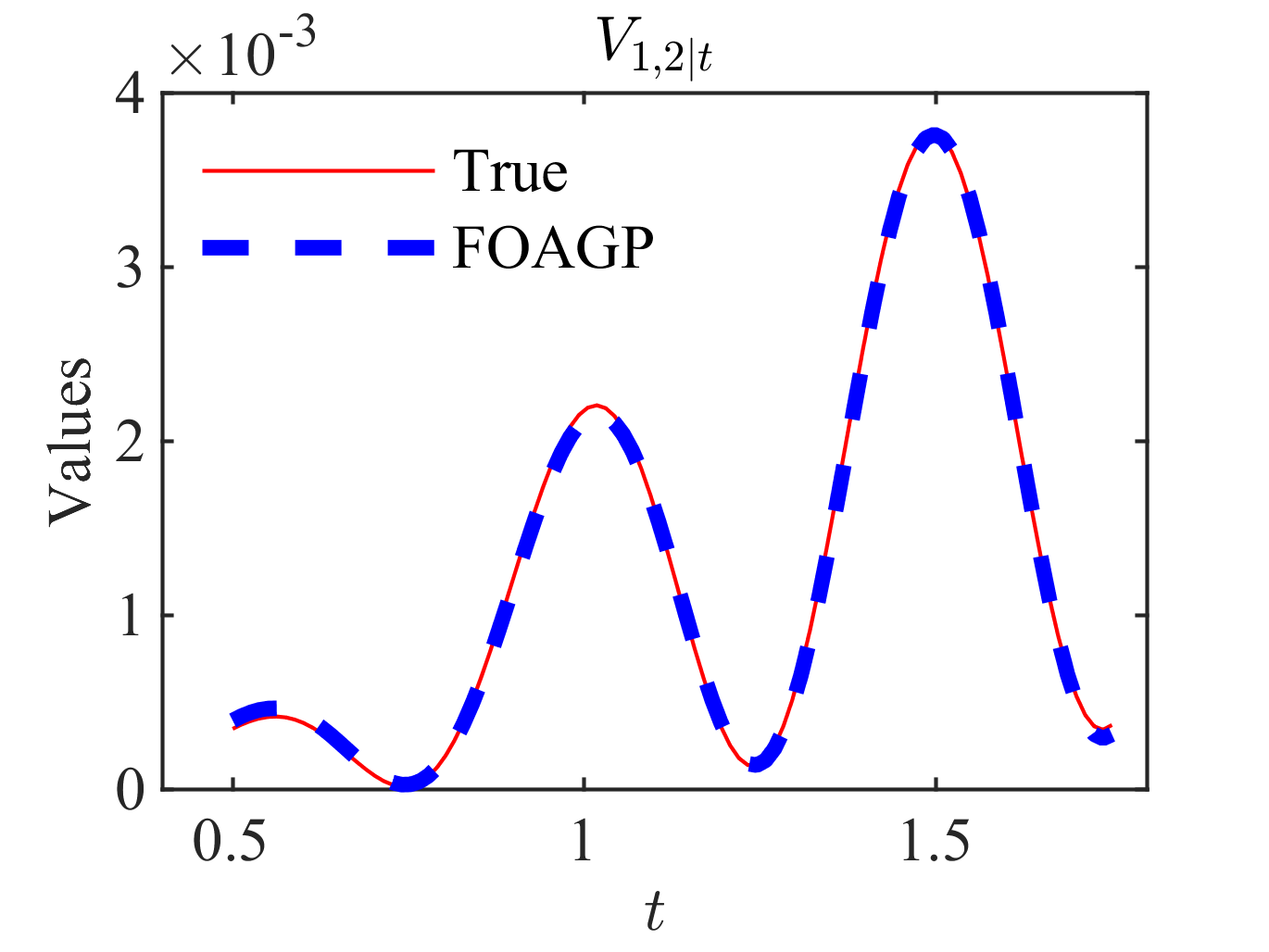}
\par\end{centering}
}
\par\end{centering}
\caption{Illustration of the local variance decomposition of the Example 2
using FOAGP. \label{fig:2010-variance}}
\end{figure*}

The local variance decomposition results using FOAGP for both examples
are shown in Fig. \ref{fig:toy-variance} and Fig. \ref{fig:2010-variance}.
These predicted results using Eq. \eqref{eq:VarEstimator} are very
close to the theoretical values, demonstrating the ability of FOAGP
to perform local variance decomposition of the functional outputs.
By leveraging the entire dataset, the data-driven estimators can accurately
estimate the local variance of the functional outputs at any output
position. This capacity distinguishes FOAGP from \citet{drigneiFunctionalANOVAComputer2010}
method, which applies FANOVA only at fixed output positions to compute
the local variance at those fixed output positions. 

It is worth noting that in Example 2, while the local variances of
the main effects are significantly higher than that of the interaction
effect, they almost vanish at some specific points. This highlights
the advantage of local variance decomposition in providing a more
refined understanding of the underlying variability, whereas relying
solely on global variance could obscure the truly important variables
at these specific points. 
 
\subsection{Global Sensitivity \textcolor{black}{Indices }}

\begin{figure*}
\begin{centering}
\subfloat[Example 1 ]{\begin{centering}
\includegraphics[width=8cm]{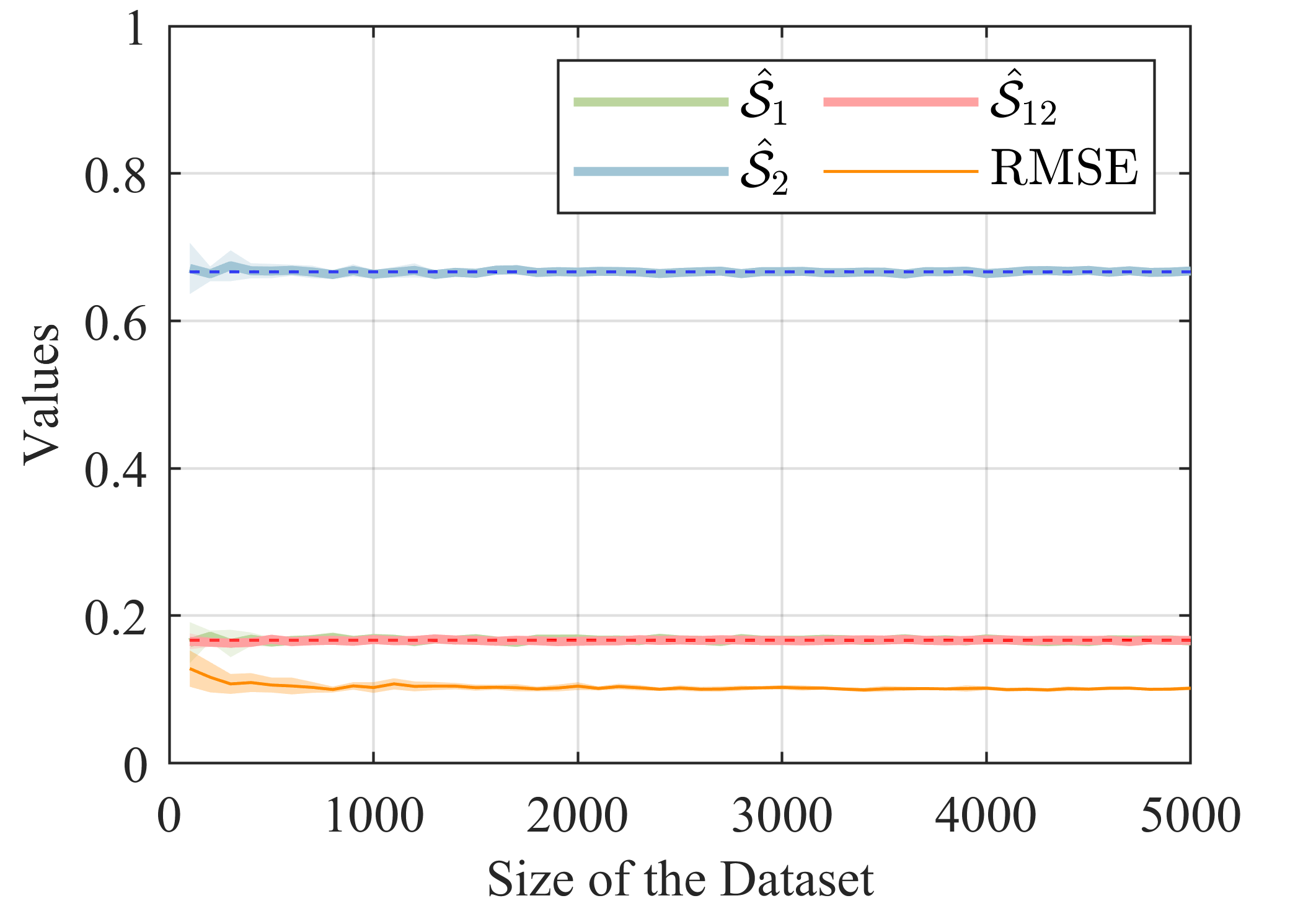}
\par\end{centering}
}\subfloat[Example 2 ]{\begin{centering}
\includegraphics[width=7.7cm]{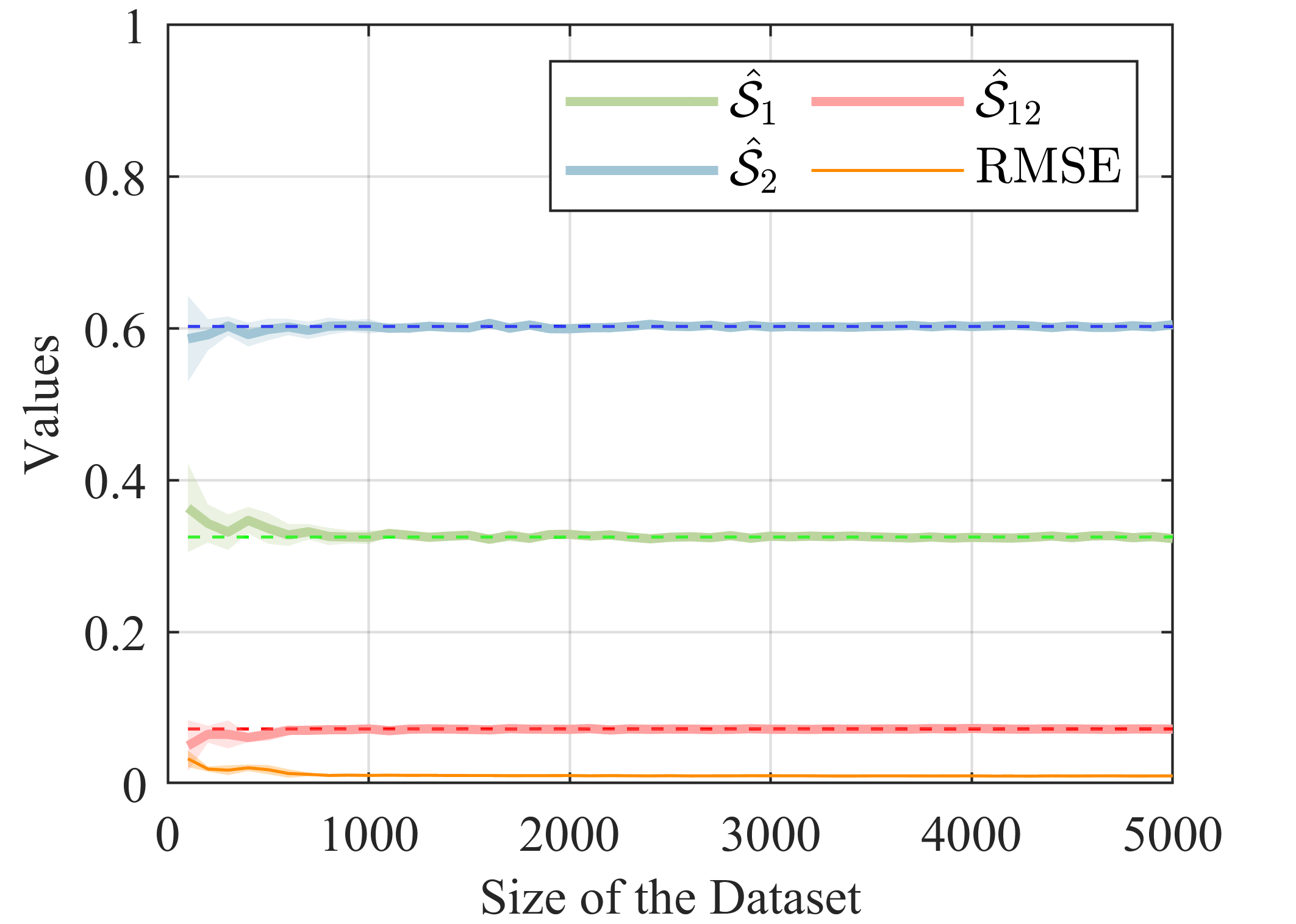}
\par\end{centering}
}
\par\end{centering}
\caption{Illustration of the estimated global sensitivity indices and RMSE
using FOAGP. Due to the theoretical ECV sensitivity indices $\mathcal{S}_{1}=\mathcal{S}_{12}$
in the example 1, the green lines are covered by the red lines in
Fig. \ref{fig:ECV}(a). \label{fig:ECV}  }
\end{figure*}

To demonstrate the effectiveness of FOAGP in estimating global sensitivity
indices, we perform simulations on two examples with varying dataset
sizes, repeating each simulation 10 times per dataset size. The dataset
sizes range from 100 to 5,000, increasing in increments of 100, with
a training-to-test ratio of 4:1. The estimated global sensitivity
indices are computed using Eq. \eqref{eq:Global Indices} and Eq.
\eqref{eq:VarEstimator}. The global sensitivity indices results using
FOAGP are shown in Fig. \ref{fig:ECV}, where the thinner dashed horizontal
lines represent the theoretical ECV sensitivity indices, the thicker
solid lines represent the estimated global sensitivity indices, the
thinner orange line represents the RMSE, and the shaded areas indicate
one standard deviation. 

In both examples, the estimated global sensitivity indices converge
rapidly to the theoretical values, empirically supporting that the
data-driven estimators in Eq. \eqref{eq:VarEstimator} are asymptotically
unbiased. Similarly, the RMSE also converges quickly in both examples,
demonstrating the ability of the nonparametric FOAGP model to accurately
capture complicated nonlinear relationships. In example 1, the estimated
global sensitivity indices closely match the theoretical values even
with very small dataset sizes. This is because the functional output
in example 1 shares the same structure as FOAGP in Eq. \eqref{eq:FOAGP}
and is relatively simple. 

\section{Real Case Study \label{sec:Real-Case-Study}  }

\begin{figure}
\begin{centering}
\includegraphics[width=8cm]{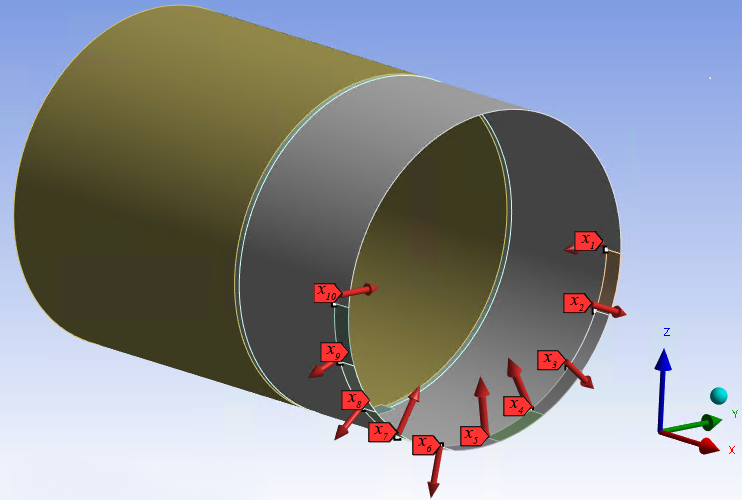}
\par\end{centering}
\caption{Composite fuselage simulation in Ansys 2024. \label{fig:Case-Study-System}}
\end{figure}

This section demonstrates the applicability of the proposed FOAGP
model with a real case study on fuselage shape control. In order to
achieve better dimensional deformation control around the edge of
the fuselage, a shape control system was developed \citep{wenFeasibilityAnalysisComposite2018}.
As shown in Fig. \ref{fig:Case-Study-System}, the system uses 10
actuators uniformly distributed along the lower half of the fuselage
in a single plane with the same X-axis, which can push or pull to
adjust the in-plane shape of the fuselage. A uniformly sampled dataset
for the case study is generated using the finite element model from
\citet{wenFeasibilityAnalysisComposite2018}. The dataset consists
of 10,000 instances, each containing 10 forces $F_{i}$ provided by
the actuators, all below 450 pound, where $F_{i}>0$ represents a
pull and $F_{i}<0$ represents a push. Additionally, 100 points are
uniformly distributed along the fuselage to measure the deformed radius.
The output for each instance represents the deformation at these 100
points, which can be viewed as a functional output of length 100.
White noise $\epsilon\sim\mathcal{N}\left(0,0.1^{2}\right)$ is added
to the outputs. 

A total of 8,000 instances are randomly selected to train the FOAGP
model, with the remaining data forming the test set. The kernel matrix
with a Kronecker product structure, as shown in Eq. \eqref{eq:KronKernel},
is utilized to accelerate parameter estimation. Additionally, corresponding
predictors and variance estimators in Eq. \eqref{eq:KronPred} and
Eq. \eqref{eq:KronVarEstimator} are employed to further improve efficiency.
To enhance numerical stability, the forces $F_{i}$ are scaled from
$\left[-450,450\right]$ to $\left[0,1\right]$ by $x_{i}=\left(F_{i}+450\right)/900$
and the angular range is scaled from $\left[0,2\pi\right)$ to $\left[0,1\right)$
by $t=\alpha/\left(2\pi\right)$. As discussed in Sec. \ref{subsec:Variance Decomposition},
these bijective transformations have no influence on the ECV sensitivity
indices. To optimize model performance, a periodic kernel $k_{t}\left(t,t^{\prime}\right)=\exp\left(-\theta_{t}^{2}\sin^{2}\left(\pi\left(t-t^{\prime}\right)/T\right)\right)$
is set for the output position $t$ \citep{mackay1998introduction},
with the periodic parameter $T=1$, while the input kernels remain
Gaussian kernels $k_{i}\left(x_{i},x_{i}^{\prime}\right)=\exp\left(-\left(x_{i}-x_{i}^{\prime}\right)^{2}/\left(2\theta_{i}^{2}\right)\right)$.
The experiments are repeated 10 times. Prior to the main experiments,
a preliminary test was conducted by incorporating the angle into the
traditional FANOVA. The resulting top five Sobol' indices are: $\phi_{t}=0.9447,\phi_{1t}=0.0155,\phi_{10,t}=0.0142,\phi_{2,t}=0.0094,\phi_{9,t}=0.0085$,
indicating that the output position $t$ overwhelmingly dominates
the variance and overshadows the contributions of the forces. 

\begin{figure}
\begin{centering}
\includegraphics[width=9cm]{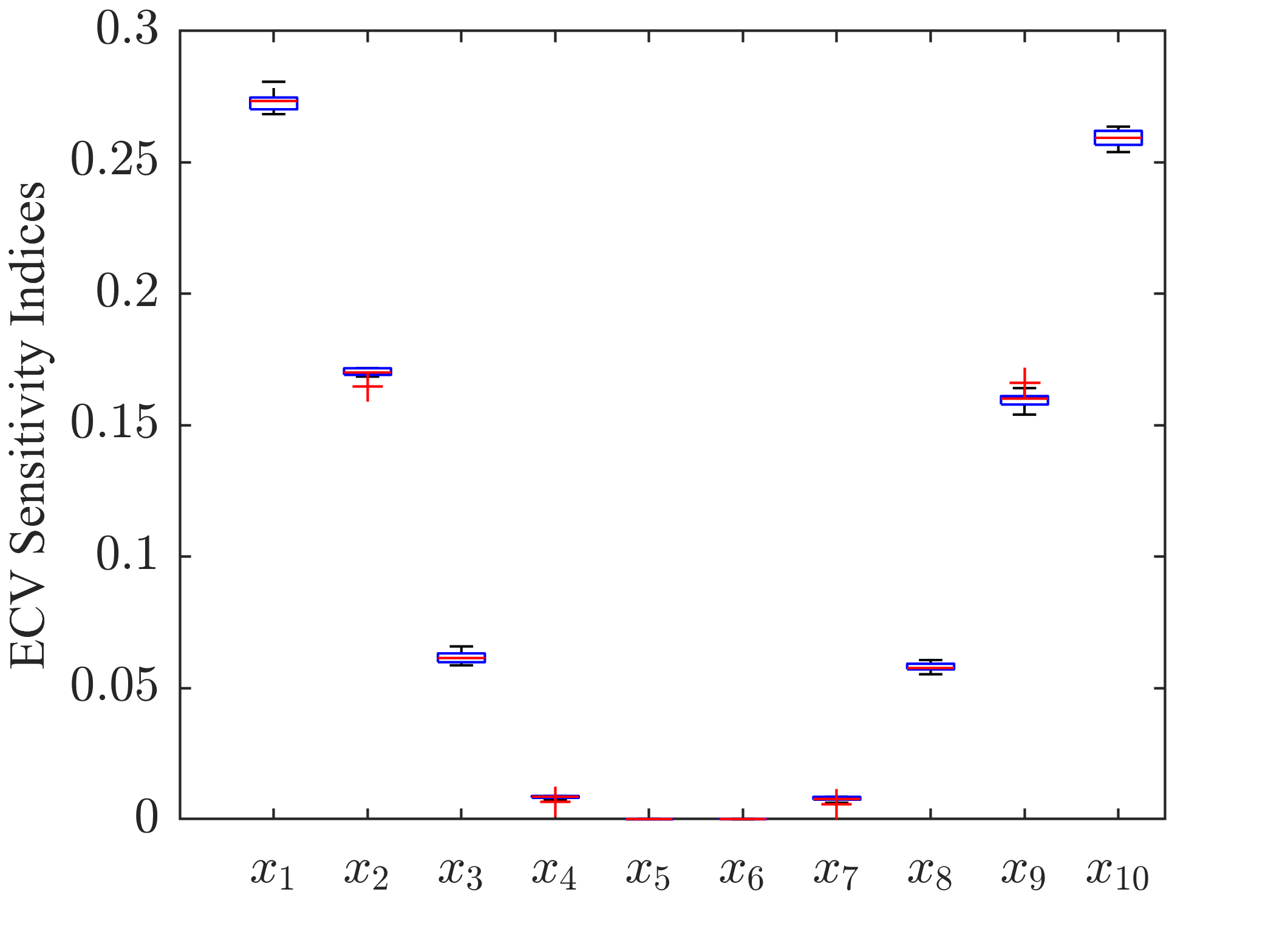}
\par\end{centering}
\caption{Box plot of the ECV sensitivity indices of the main effects. \label{fig:Case-Study-ECV-Boxplot}}
\end{figure}

Fig. \ref{fig:Case-Study-ECV-Boxplot} shows the box plot of the ECV
sensitivity indices for each main effect, with cumulative ECV sensitivity
indices of the main effects exceed \textcolor{black}{0.9977} across
all experiments, which indicates that the main effects dominate the
variance. This is consistent with the fact that the actuators are
positioned within the same plane, where the forces behave as vectors.
Consequently, the resulting deformation should approximately follow
vector arithmetic rules, with negligible interaction effects. In addition,
the actuators located farther from lowest point have a greater influence,
while those near the lowest point have minimal influence. 
 
\begin{figure}
\begin{centering}
\subfloat[]{\begin{centering}
\includegraphics[width=3cm]{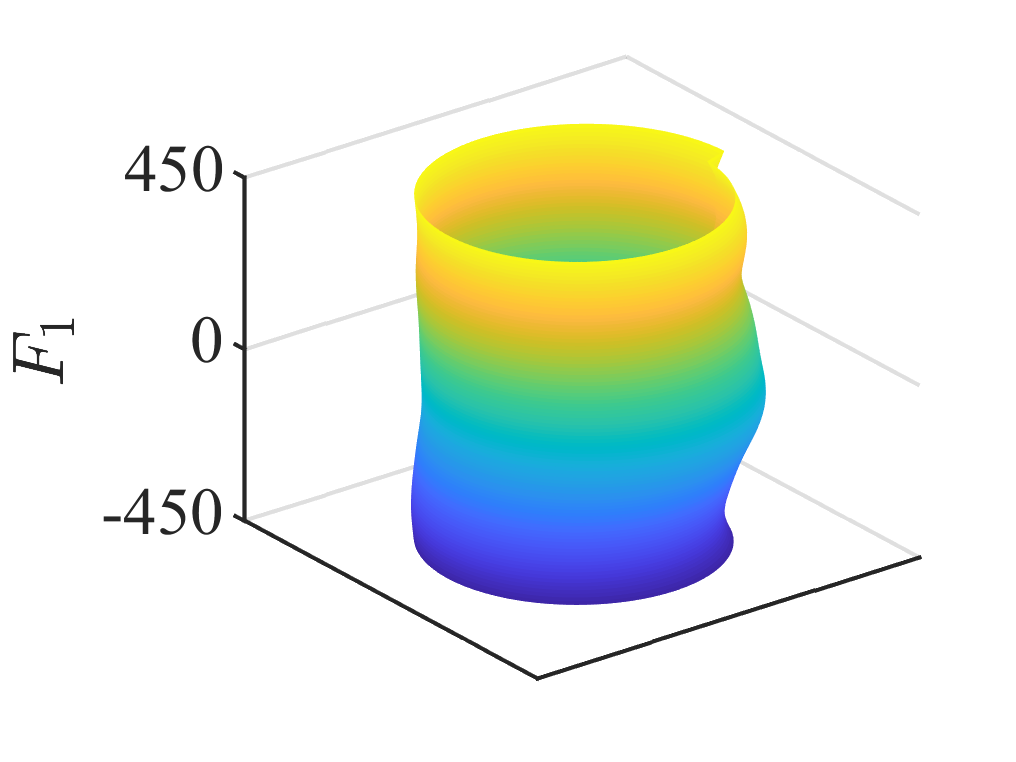}
\par\end{centering}
}\subfloat[]{\begin{centering}
\includegraphics[width=3cm]{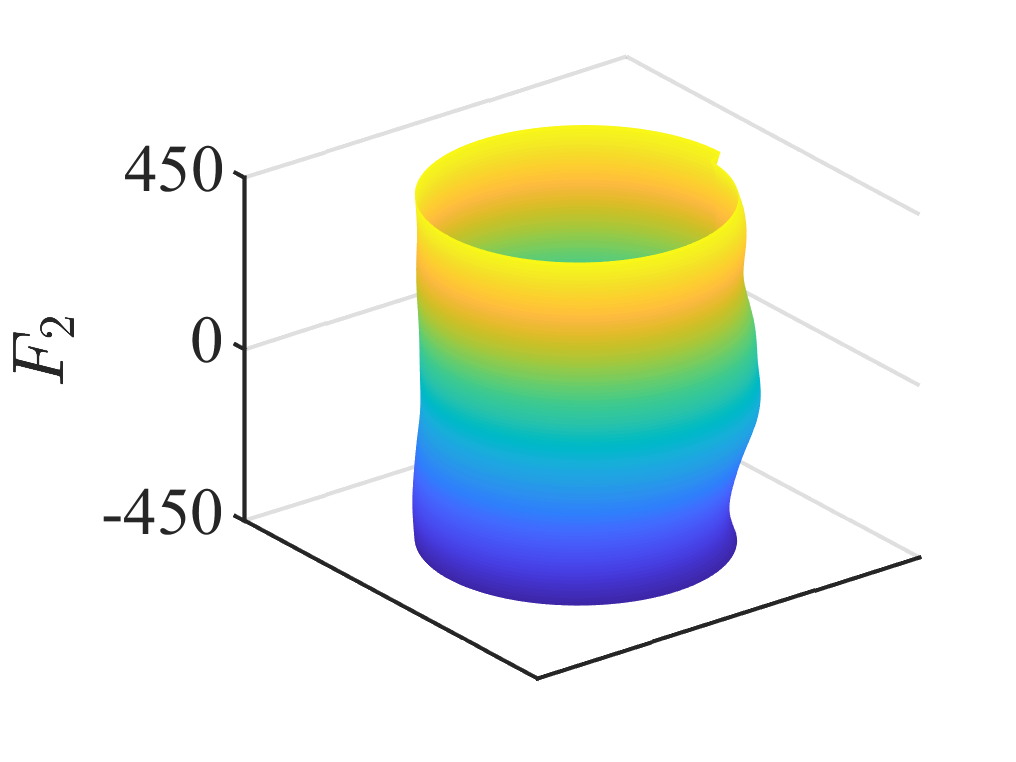}
\par\end{centering}
}\subfloat[]{\begin{centering}
\includegraphics[width=3cm]{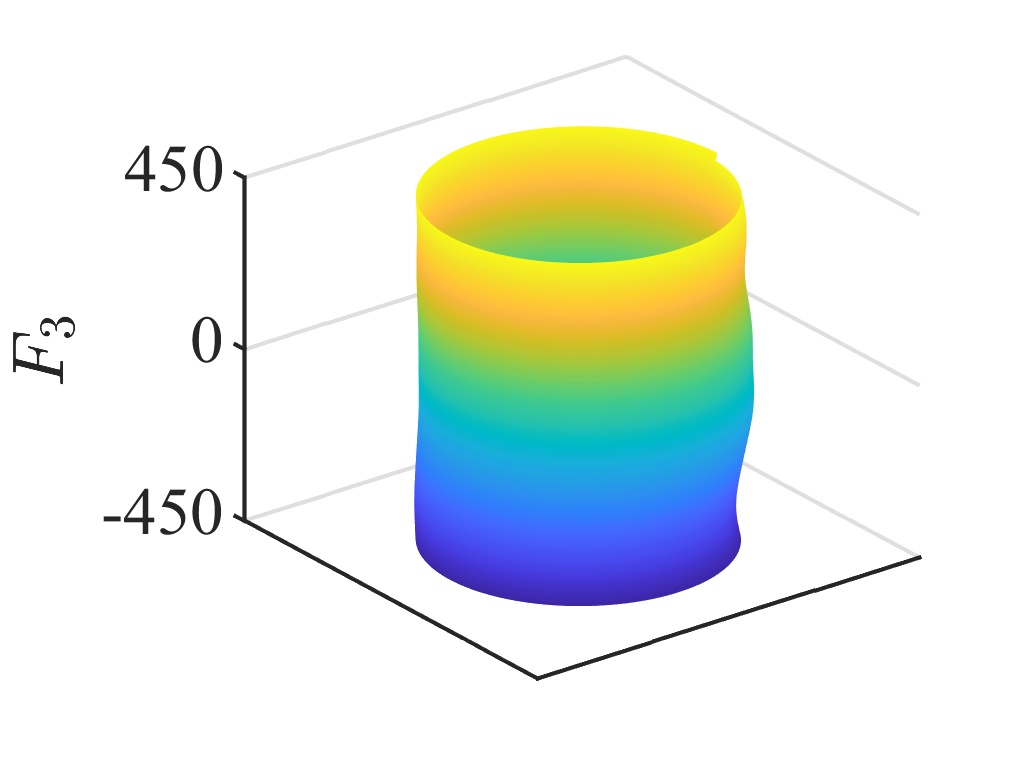}
\par\end{centering}
}\subfloat[]{\begin{centering}
\includegraphics[width=3cm]{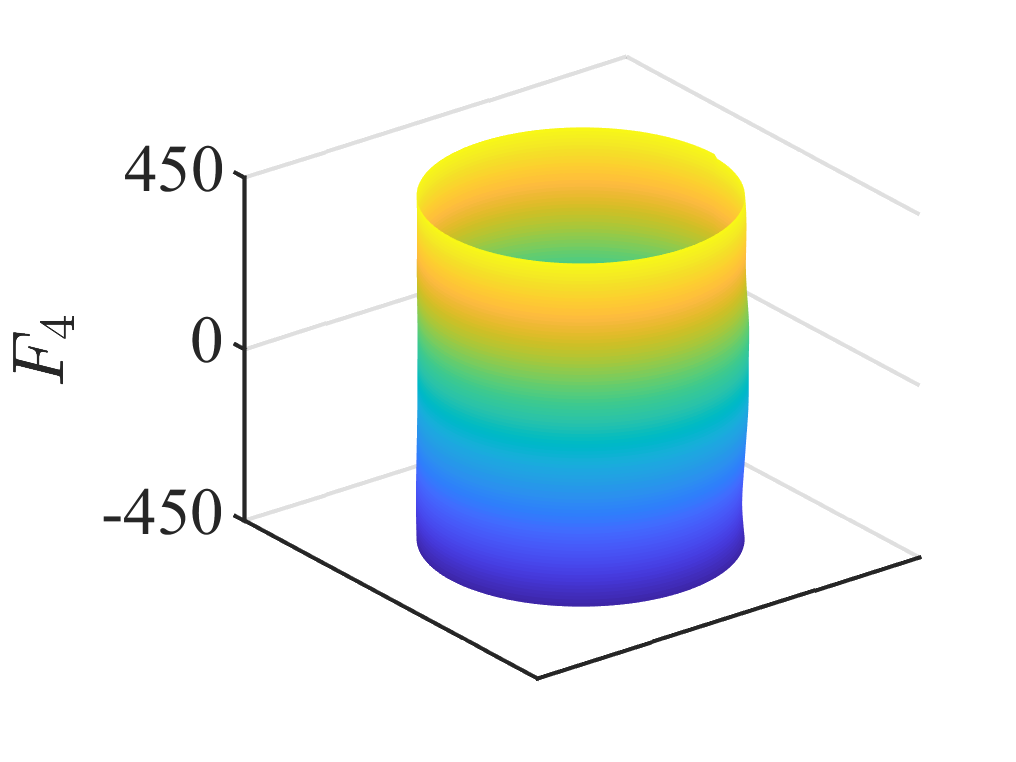}
\par\end{centering}
}\subfloat[]{\begin{centering}
\includegraphics[width=3cm]{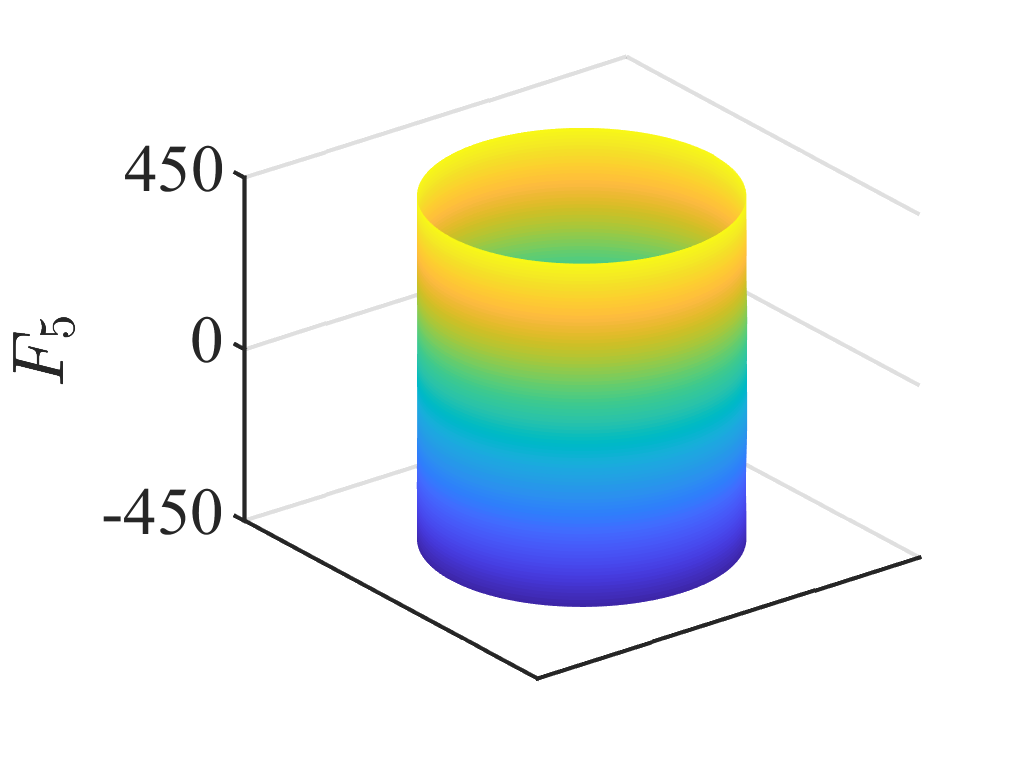}
\par\end{centering}
}
\par\end{centering}
\begin{centering}
\subfloat[]{\begin{centering}
\includegraphics[width=3cm]{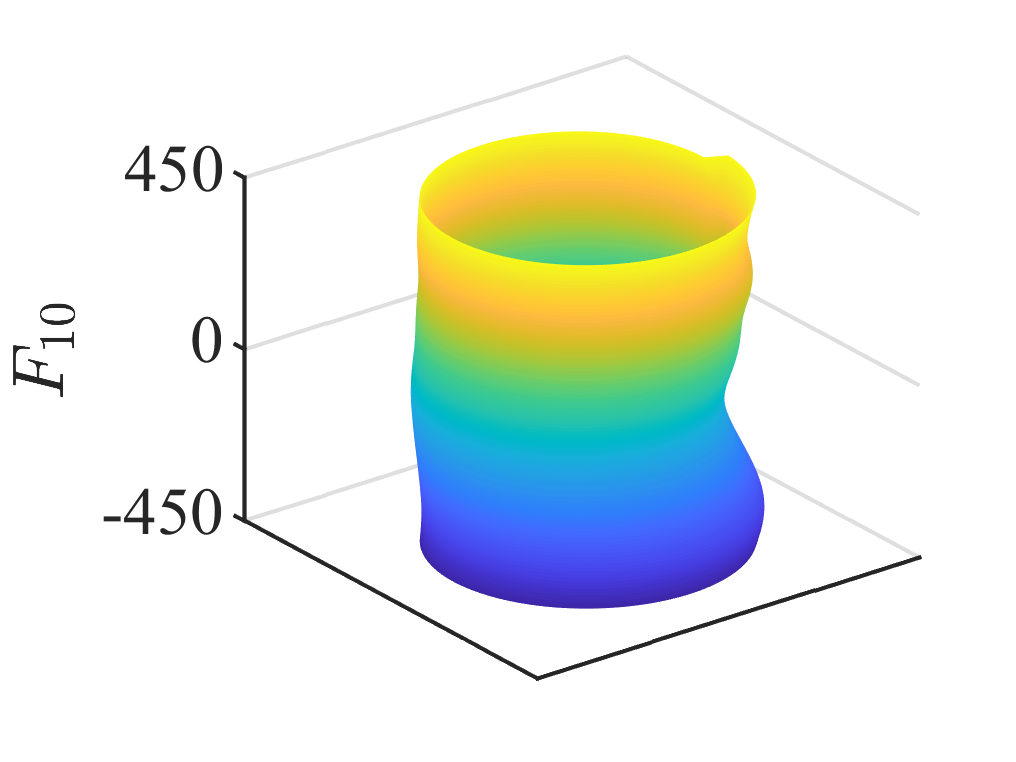}
\par\end{centering}
}\subfloat[]{\begin{centering}
\includegraphics[width=3cm]{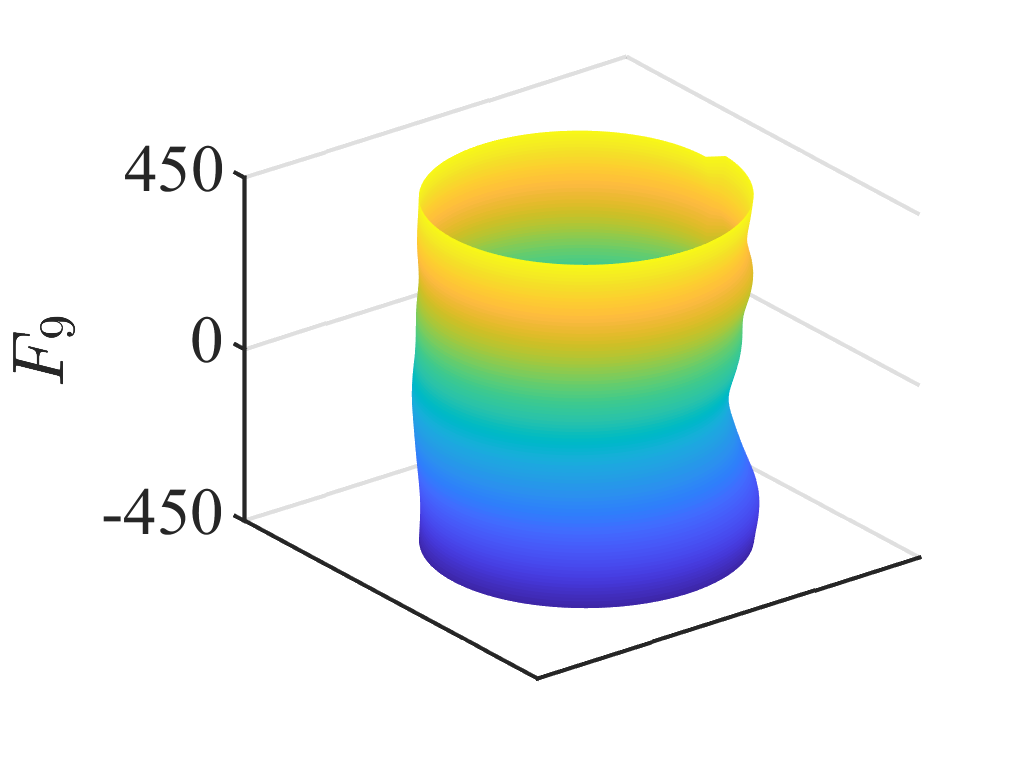}
\par\end{centering}
}\subfloat[]{\begin{centering}
\includegraphics[width=3cm]{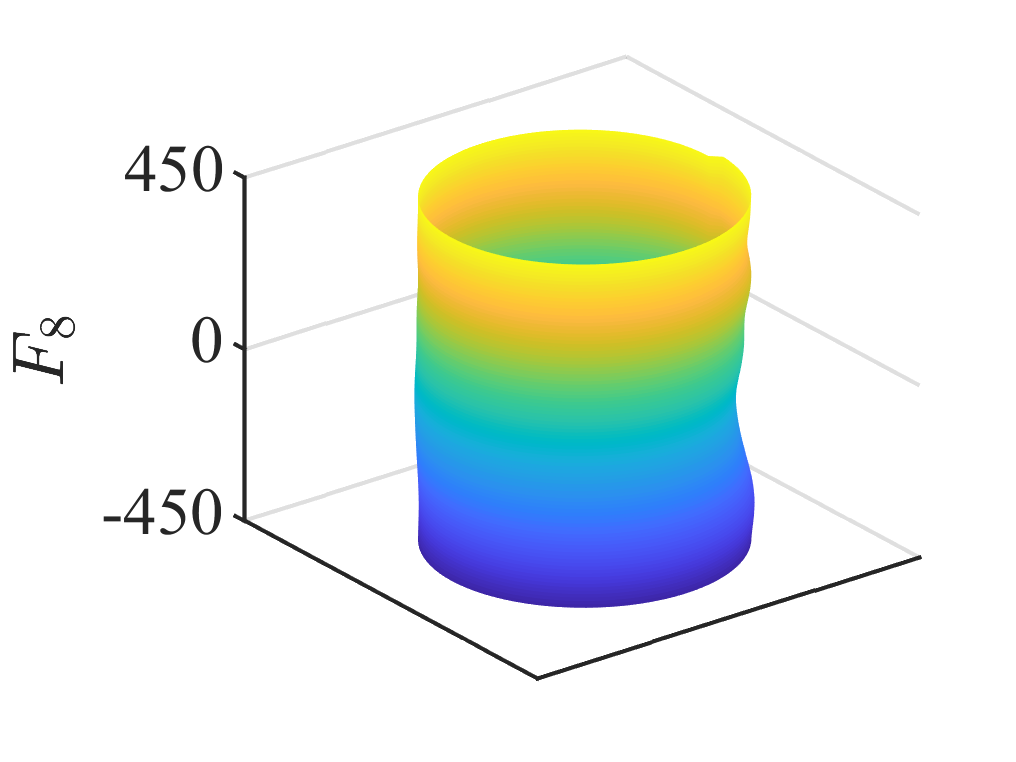}
\par\end{centering}
}\subfloat[]{\begin{centering}
\includegraphics[width=3cm]{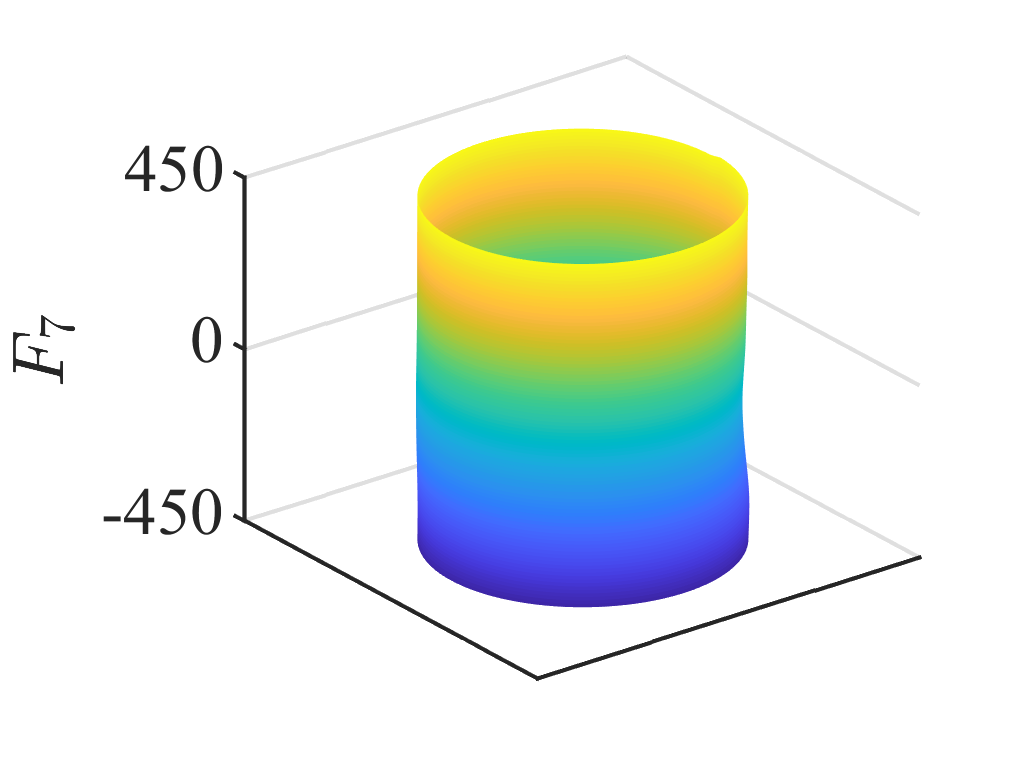}
\par\end{centering}
}\subfloat[]{\begin{centering}
\includegraphics[width=3cm]{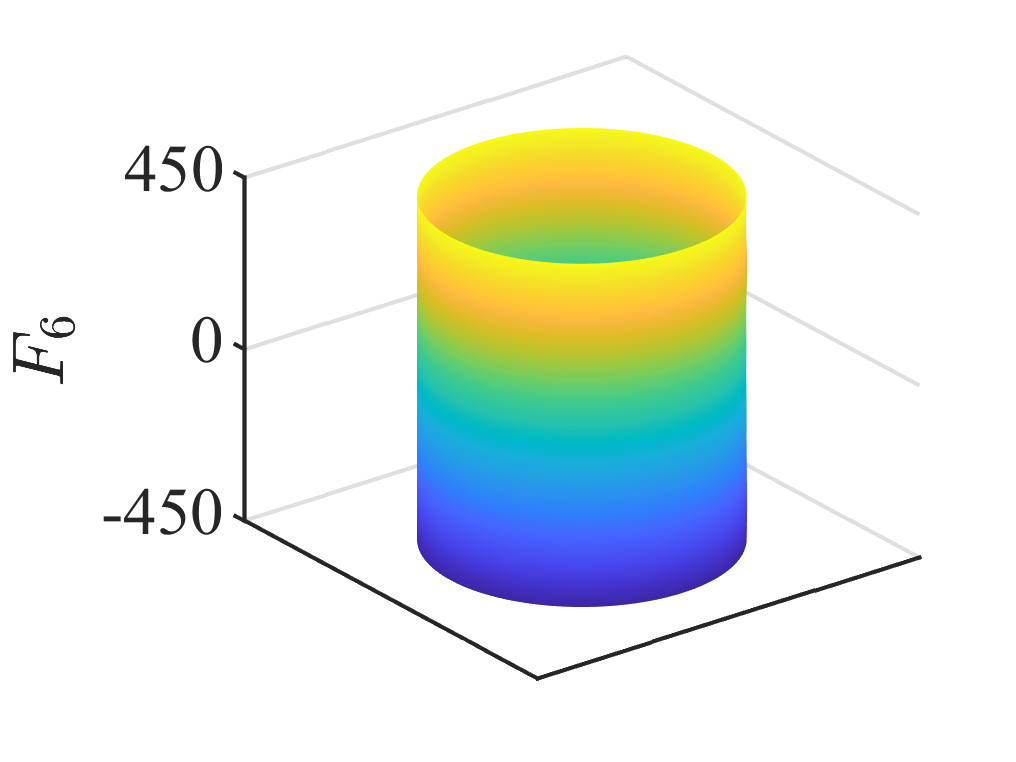}
\par\end{centering}
}
\par\end{centering}
\caption{The deformation cased by the main effects of the forces. To illustrate
the deformation better, we scale the original radius to 1/10 and the
deformation remains unchanged. \label{fig:Case-Study-Deformation Visual}
}
\end{figure}

Fig. \ref{fig:Case-Study-Deformation Visual} visualizes the deformation
caused by the main effects of the actuators. The edge curves illustrate
the main effects of forces on the fuselage shape. Actuators located
farther from the lowest point, particularly the 1st and 10th actuators,
exert greater and more symmetric influences on the fuselage shape
due to their mid-height placement. This suggests that focusing on
these key actuators can improve control efficiency and system reliability.
Conversely, the minimal impact of actuators near the lowest point
indicates potential opportunities for system simplification and cost
reduction without compromising performance. This understanding enables
engineers to prioritize actuator placement strategically, optimizing
both the performance and cost-effectiveness in fuselage shape control
systems. These results demonstrate the capability of FOAGP to perform
orthogonal FOED and sensitivity analysis in real-world scenarios,
offering a unified data-driven implementation of FOFANOVA with significant
practical value. 

\section{Conclusions \label{sec:Conclusions}}

In this study, we propose the FOAGP model, a novel data-driven, nonlinear,
and nonparametric modeling approach for efficient FOED. The FOAGP
model fully incorporates high-order effects and embeds FOED directly
into model structure, reducing the time cost associated with separate
decomposition steps. Utilizing the proposed functional-output orthogonal
additive kernel, FOAGP achieves orthogonal FOED without requiring
prior knowledge about the underlying data distribution. Furthermore,
a corresponding global local variance decomposition is established
within the FOAGP framework, enabling comprehensive global local sensitivity
analysis for functional-output computer experiments. The effectiveness
of FOAGP is demonstrated through two simulation studies and a real-world
case study on fuselage shape control. These experiments validate the
model's capability to perform FOED and accurately quantify variable
importance, highlighting FOAGP as a practical and handy tool for FOFANOVA
in functional-output computer experiments, and demonstrating its potential
values in complex engineering scenarios. 

While this study focuses only on one-dimensional functional outputs,
the method can be readily extendable to high-dimensional functional
output case, which will be explored in the future research. The proposed
FOAGP model is under the assumption of no input-output interactions.
In scenarios where such interactions exist, the model assumptions
may need to be revised. Future work will explore incorporating input-output
interactions through frameworks such as generalized FANOVA and Shapley
values. Another promising direction involves leveraging functional-output
effect decomposition within the context of Bayesian functional optimization,
which may open new avenues for efficient optimization in functional
spaces. 

\if0\blind { 

\section*{Acknowledgments}

We would like to express our sincere gratitude to Jeff Wu and Jianhua
Huang for their fruitful comments for this paper.

} \fi 

\bigskip{}

\begin{spacing}{1}\bibliographystyle{asa}
\bibliography{AGP}

\end{spacing}
\clearpage
\noindent \begin{center}
\textbf{\LARGE{}Supplemental Material to ``Effect Decomposition of
Functional-Output Computer Experiments via Orthogonal Additive Gaussian
Processes''}{\LARGE\par}
\par\end{center}

\noindent \bigskip{}

\setcounter{section}{0}

\section{Proof}

\begin{proofTheorem}Without loss of generality, we assume that the
input variable $x_{i}$ and the time $t$ both lie in the interval
$\left[0,1\right]$ with a uniform distribution. If the range of $x_{i}$
or $t$ is not within $\left[0,1\right]$, a bijective transformation
can be applied to map the range to $\left[0,1\right]$. If the distribution
of $x_{i}$ or $t$ are not uniform, the differential $\mathrm{d}x_{i}$
or $\mathrm{d}t$ used in the following proof can be replaced by the
measure $\mathrm{d}F_{i}\left(x_{i}\right)$ or $\mathrm{d}F_{t}\left(t\right)$,
respectively. 

Given the separable prior process
\[
f\sim\mathcal{G}\mathcal{P}\left(\mathbf{0},k_{i}k_{t}\right),
\]
where $f:\left[0,1\right]^{2}\rightarrow\mathbb{R}$. Under the grid
design on the input space $\left\{ \left(x_{i,u}^{\left(n\right)},t_{v}^{\left(n\right)}\right)\right\} _{u,v=1}^{n}\subseteq\left[0,1\right]^{2}$,
where $x_{i,u}^{\left(n\right)}=u/n$ and $t_{v}^{\left(n\right)}=v/n$,
the orthogonality constraint can be approximated by 
\[
z_{v}^{\left(n\right)}=\int f\left(x_{i},t_{v}^{\left(n\right)}\right)\mathrm{d}x_{i}\approx\frac{1}{n}\sum_{u=1}^{n}f\left(x_{i,u}^{\left(n\right)},t_{v}^{\left(n\right)}\right)=\boldsymbol{a}_{n}^{\top}\boldsymbol{f}_{v}^{\left(n\right)},
\]
where $\boldsymbol{a}_{n}$ is a $n\times1$ vector comprising all
the elements equal to $1/n$, $\boldsymbol{f}_{v}^{\left(n\right)}=\left[f\left(x_{i,1}^{\left(n\right)},t_{v}^{\left(n\right)}\right),\cdots,\left(x_{i,n}^{\left(n\right)},t_{v}^{\left(n\right)}\right)\right]^{\top}$.
Let 
\[
\boldsymbol{z}^{\left(n\right)}=\left[\begin{array}{c}
z_{1}^{\left(n\right)}\\
\vdots\\
z_{n}^{\left(n\right)}
\end{array}\right]=\left(\boldsymbol{I}_{n}\otimes\boldsymbol{a}_{n}\right)^{\top}\left[\begin{array}{c}
\boldsymbol{f}_{1}^{\left(n\right)}\\
\vdots\\
\boldsymbol{f}_{n}^{\left(n\right)}
\end{array}\right]=\left(\boldsymbol{I}_{n}\otimes\boldsymbol{a}_{n}\right)^{\top}\boldsymbol{f}^{\left(n\right)},
\]
where $\boldsymbol{f}^{\left(n\right)}=\left[\begin{matrix}\left(\boldsymbol{f}_{1}^{\left(n\right)}\right)^{\top} & \cdots & \left(\boldsymbol{f}_{n}^{\left(n\right)}\right)^{\top}\end{matrix}\right]^{\top}$.
It is easy to show that $\mathrm{\mathbb{V}}\left(\boldsymbol{f}^{\left(n\right)}\right)=\boldsymbol{K}_{t}^{\left(n\right)}\otimes\boldsymbol{K}_{i}^{\left(n\right)}$,
$\mathrm{\mathbb{V}}\left(\boldsymbol{z}^{\left(n\right)}\right)=\boldsymbol{K}_{t}^{\left(n\right)}\otimes\left(\boldsymbol{a}_{n}^{\top}\boldsymbol{K}_{i}^{\left(n\right)}\boldsymbol{a}_{n}\right)$
and $\text{Cov}\left(\boldsymbol{f}^{\left(n\right)},\boldsymbol{z}^{\left(n\right)}\right)=\boldsymbol{K}_{t}^{\left(n\right)}\otimes\left(\boldsymbol{K}_{i}^{\left(n\right)}\boldsymbol{a}_{n}\right)$,
where $\boldsymbol{K}_{t}^{\left(n\right)}=\left[k_{t}\left(t_{u}^{\left(n\right)},t_{v}^{\left(n\right)}\right)\right]_{u,v=1,\cdots,n}$
and $\boldsymbol{K}_{i}^{\left(n\right)}=\left[k_{i}\left(x_{i,u}^{\left(n\right)},x_{i,v}^{\left(n\right)}\right)\right]_{u,v=1,\cdots,n}$.
By the the definition of Gaussian Process, the joint distribution
of\textcolor{teal}{{} }$\left(\boldsymbol{f}^{\left(n\right)},\boldsymbol{z}^{\left(n\right)}\right)$
is Gaussian, and it is easy to show that 
\[
\left[\begin{array}{c}
\boldsymbol{f}^{\left(n\right)}\\
\boldsymbol{z}^{\left(n\right)}
\end{array}\right]\sim\left(\boldsymbol{0},\left[\begin{matrix}\boldsymbol{K}_{t}^{\left(n\right)}\otimes\boldsymbol{K}_{i}^{\left(n\right)} & \boldsymbol{K}_{t}^{\left(n\right)}\otimes\left(\boldsymbol{K}_{i}^{\left(n\right)}\boldsymbol{a}_{n}\right)\\
\boldsymbol{K}_{t}^{\left(n\right)}\otimes\left(\boldsymbol{a}_{n}^{\top}\boldsymbol{K}_{i}^{\left(n\right)}\right) & \boldsymbol{K}_{t}^{\left(n\right)}\otimes\left(\boldsymbol{a}_{n}^{\top}\boldsymbol{K}_{i}^{\left(n\right)}\boldsymbol{a}_{n}\right)
\end{matrix}\right]\right).
\]
So the posterior distribution of $\boldsymbol{f}^{\left(n\right)}\mid\left\{ \boldsymbol{z}^{\left(n\right)}=\mathbf{0}\right\} $
can be derived 
\begin{align}
 & \boldsymbol{f}^{\left(n\right)}\mid\left\{ \boldsymbol{z}^{\left(n\right)}=\mathbf{0}\right\} \nonumber \\
\sim & \mathcal{N}\left(\mathbf{0},\boldsymbol{K}_{t}^{\left(n\right)}\otimes\boldsymbol{K}_{i}^{\left(n\right)}-\left(\boldsymbol{K}_{t}^{\left(n\right)}\otimes\left(\boldsymbol{K}_{i}^{\left(n\right)}\boldsymbol{a}_{n}\right)\right)\left(\boldsymbol{K}_{t}^{\left(n\right)}\otimes\left(\boldsymbol{a}_{n}^{\top}\boldsymbol{K}_{i}^{\left(n\right)}\boldsymbol{a}_{n}\right)\right)^{-1}\left(\boldsymbol{K}_{t}^{\left(n\right)}\otimes\left(\boldsymbol{K}_{i}^{\left(n\right)}\boldsymbol{a}_{n}\right)^{\top}\right)\right)\nonumber \\
= & \mathcal{N}\left(\mathbf{0},\boldsymbol{K}_{t}^{\left(n\right)}\otimes\boldsymbol{K}_{i}^{\left(n\right)}-\boldsymbol{K}_{t}^{\left(n\right)}\otimes\frac{\left(\boldsymbol{K}_{i}^{\left(n\right)}\boldsymbol{a}_{n}\right)\left(\boldsymbol{K}_{i}^{\left(n\right)}\boldsymbol{a}_{n}\right)^{\top}}{\boldsymbol{a}_{n}^{\top}\boldsymbol{K}_{i}^{\left(n\right)}\boldsymbol{a}_{n}}\right)\nonumber \\
= & \mathcal{N}\left(\mathbf{0},\boldsymbol{K}_{t}^{\left(n\right)}\otimes\left(\boldsymbol{K}_{i}^{\left(n\right)}-\frac{\left(\boldsymbol{K}_{i}^{\left(n\right)}\boldsymbol{a}_{n}\right)\left(\boldsymbol{K}_{i}^{\left(n\right)}\boldsymbol{a}_{n}\right)^{\top}}{\boldsymbol{a}_{n}^{\top}\boldsymbol{K}_{i}^{\left(n\right)}\boldsymbol{a}_{n}}\right)\right).\label{eq:posterior dis}
\end{align}
For any two points $\left(x_{i},t\right)$ and $\left(x_{i}^{\prime},t^{\prime}\right)$,
there exist two sequences $\left(x_{\alpha}^{\left(n\right)},t_{\xi}^{\left(n\right)}\right)$
and $\left(x_{\beta}^{\left(n\right)},t_{\eta}^{\left(n\right)}\right)$
in $\left[0,1\right]\times\left[0,1\right]$ such that when $n$ tends
to infinity 
\[
\left(x_{\alpha}^{\left(n\right)},t_{\xi}^{\left(n\right)}\right)\rightarrow\left(x_{i},t\right),\text{ and }\left(x_{\beta}^{\left(n\right)},t_{\eta}^{\left(n\right)}\right)\rightarrow\left(x_{i}^{\prime},t^{\prime}\right).
\]
It should be noted that here $\alpha$, $\beta$, $\xi$, and $\eta$
are dependent on $n$. The correlation between $\left(x_{\alpha}^{\left(n\right)},t_{\xi}^{\left(n\right)}\right)$
and $\left(x_{\beta}^{\left(n\right)},t_{\eta}^{\left(n\right)}\right)$
is exactly the $\left(\xi-1\right)n+\alpha$th row and $\left(\eta-1\right)n+\beta$th
column of the kernel matrix in Eq. \eqref{eq:posterior dis}, i.e.,
\[
k_{t}\left(t_{\xi}^{\left(n\right)},t_{\eta}^{\left(n\right)}\right)\left(k\left(x_{\alpha}^{\left(n\right)},x_{\beta}^{\left(n\right)}\right)-\frac{\left[\frac{1}{n}\sum_{u=1}^{n}k\left(x_{i,u},x_{\alpha}^{\left(n\right)}\right)\right]\left[\frac{1}{n}\sum_{v=1}^{n}k\left(x_{i,v},x_{\beta}^{\left(n\right)}\right)\right]}{\frac{1}{n^{2}}\sum_{u,v=1}^{n}k\left(x_{i,u},x_{i,v}\right)}\right).
\]
Taking the limit of $n\to\infty$ yields 
\begin{align*}
\tilde{k}_{i\mid t}\left(\left(x_{i},t\right),\left(x_{i}^{\prime},t^{\prime}\right)\right) & =k_{t}\left(t,t^{\prime}\right)\left(k\left(x_{i},x_{i}^{\prime}\right)-\frac{\mathbb{E}_{x}\left[k\left(x,x_{i}\right)\right]\mathbb{E}_{x^{\prime}}\left[k\left(x^{\prime},x_{i}^{\prime}\right)\right]}{\mathbb{E}_{x,x^{\prime}}\left[k\left(x,x^{\prime}\right)\right]}\right)\\
 & =k_{t}\left(t,t^{\prime}\right)\tilde{k}\left(x_{i},x_{i}^{\prime}\right),
\end{align*}
where $\tilde{k}\left(x_{i},x_{i}^{\prime}\right)$ is exactly the
orthogonal  kernel. Taking the limit of $n\to\infty$ in Eq. \eqref{eq:posterior dis}
gives the conditional GP  
\[
f_{i\mid t}\mid\left\{ \int f_{i\mid t}\left(x_{i},t\right)\mathrm{d}F_{i}\left(x_{i}\right)=0\right\} \sim\mathcal{G}\mathcal{P}\left(0,\tilde{k}_{i}k_{t}\right),
\]
which finishes the proof. \end{proofTheorem}

\begin{proofTheorem} The representer theorem \citep{kimeldorfResultsTchebycheffianSpline1971}
shows that the effect prediction $\hat{f}_{\boldsymbol{u}\mid t}\left(\boldsymbol{x}_{\boldsymbol{u}},t\right)$
lies within the span of the corresponding  kernel $\tilde{k}_{\boldsymbol{u}\mid t}\left(\cdot\right)$.
It suffices to show that the conditional zero mean and conditional
orthogonality hold for the proposed kernels $\left\{ \tilde{k}_{\boldsymbol{u}\mid t}\left(\cdot\right)\right\} $.
By the construction of the orthogonal kernels $\left\{ \tilde{k}_{i}\left(\cdot\right)\right\} $
in Theorem 1, it is straightforward to verify that $\mathbb{E}_{x_{i}}\left[\tilde{k}_{i}\left(x_{i},x_{i}^{\prime}\right)\right]=0$. 

(a) Conditional zero mean: For any fixed point $\left(\boldsymbol{x}^{\prime},t^{\prime}\right)$,
we have 
\begin{align*}
\int k_{\boldsymbol{u}\mid t}\left(\left(\boldsymbol{x}_{\boldsymbol{u}},t\right),\left(\boldsymbol{x}_{\boldsymbol{u}}^{\prime},t^{\prime}\right)\right)\mathrm{d}F\left(\boldsymbol{x}\right) & =k_{t}\left(t,t^{\prime}\right)\int\prod_{i\in\boldsymbol{u}}\tilde{k}_{i}\left(x_{i},x_{i}^{\prime}\right)\mathrm{d}F\left(\boldsymbol{x}\right)\\
 & =k_{t}\left(t,t^{\prime}\right)\prod_{i\in\boldsymbol{u}}\int\tilde{k}_{i}\left(x_{i},x_{i}^{\prime}\right)\mathrm{d}F_{i}\left(x_{i}\right)\\
 & =k_{t}\left(t,t^{\prime}\right)\prod_{i\in\boldsymbol{u}}\mathbb{E}_{x_{i}}\left[\tilde{k}_{i}\left(x_{i},x_{i}^{\prime}\right)\right]\\
 & =0,
\end{align*}
where the first equality follows from the definition of the proposed
kernel $\tilde{k}_{\boldsymbol{u}\mid t}\left(\cdot\right)$, the
second equality leverages the independence of $\left\{ x_{i}\right\} $,
the third equality uses the definition of expectation, and the last
equality is based on the previous result $\mathbb{E}_{x_{i}}\left[\tilde{k}_{i}\left(x_{i},x_{i}^{\prime}\right)\right]=0$. 

(b) Conditional orthogonality: When either $\boldsymbol{u}$ or $\boldsymbol{v}$
is the null set $\varnothing$, the conditional orthogonality degenerates
to the conditional mean. In all other cases, where $\boldsymbol{u}\ne\boldsymbol{v}$,
there exists $i^{\ast}\in\left\{ 1,\cdots,d\right\} $ such that $i^{\ast}\in\boldsymbol{u}\Delta\boldsymbol{v}=\left(\boldsymbol{u}\setminus\boldsymbol{v}\right)\cup\left(\boldsymbol{v}\setminus\boldsymbol{u}\right)$,
where $\setminus$ denotes the set difference and $\Delta$ denotes
the symmetric difference, meaning $\boldsymbol{u}\Delta\boldsymbol{v}\ne\varnothing$.
For any fixed point $\left(\boldsymbol{x}^{\prime},t^{\prime}\right)$,
we have 
\begin{align*}
 & \int k_{\boldsymbol{u}\mid t}\left(\left(\boldsymbol{x}_{\boldsymbol{u}},t\right),\left(\boldsymbol{x}_{\boldsymbol{u}}^{\prime},t^{\prime}\right)\right)k_{\boldsymbol{v}\mid t}\left(\left(\boldsymbol{x}_{\boldsymbol{v}},t\right),\left(\boldsymbol{x}_{\boldsymbol{v}}^{\prime},t^{\prime}\right)\right)\mathrm{d}F\left(\boldsymbol{x}\right)\\
= & k_{t}\left(t,t^{\prime}\right)\int\left(\prod_{i\in\boldsymbol{u}\cap\boldsymbol{v}}\left[\tilde{k}_{i}\left(x_{i},x_{i}^{\prime}\right)\right]^{2}\right)\left(\prod_{i\in\boldsymbol{u}\Delta\boldsymbol{v}}\tilde{k}_{i}\left(x_{i},x_{i}^{\prime}\right)\right)\prod_{i\in\boldsymbol{u}\cup\boldsymbol{v}}\mathrm{d}F_{i}\left(x_{i}\right)\\
= & k_{t}\left(t,t^{\prime}\right)\left(\prod_{i\in\boldsymbol{u}\cap\boldsymbol{v}}\int\left[\tilde{k}_{i}\left(x_{i},x_{i}^{\prime}\right)\right]^{2}\mathrm{d}F\left(x_{i}\right)\right)\prod_{i\in\boldsymbol{u}\Delta\boldsymbol{v}}\int\tilde{k}_{i}\left(x_{i},x_{i}^{\prime}\right)\mathrm{d}F_{i}\left(x_{i}\right)\\
= & k_{t}\left(t,t^{\prime}\right)\left(\prod_{i\in\boldsymbol{u}\cap\boldsymbol{v}}\int\left[\tilde{k}_{i}\left(x_{i},x_{i}^{\prime}\right)\right]^{2}\mathrm{d}F\left(x_{i}\right)\right)\prod_{i\in\boldsymbol{u}\Delta\boldsymbol{v}}\mathbb{E}_{x_{i}}\left[\tilde{k}_{i}\left(x_{i},x_{i}^{\prime}\right)\right]\\
= & 0.
\end{align*}
The first equality follows from the definition of proposed kernel
$\tilde{k}_{\boldsymbol{u}\mid t}\left(\cdot\right)$, the identity
$\boldsymbol{u}\cup\boldsymbol{v}=\left(\boldsymbol{u}\cap\boldsymbol{v}\right)\sqcup\left(\boldsymbol{u}\Delta\boldsymbol{v}\right)$,
where $\sqcup$ denotes the disjoint union, the second equality leverages
the independence of $\left\{ x_{i}\right\} $, the third equality
uses the definition of expectation, and the last equality is based
on the previous result $\mathbb{E}_{x_{i}}\left[\tilde{k}_{i}\left(x_{i},x_{i}^{\prime}\right)\right]=0$.
This finishes the proof. \end{proofTheorem}

\begin{proofTheorem} The conditional zero mean property and the conditional
orthogonality in Theorem 2 imply that 
\begin{equation}
\mathrm{Cov}\left(\hat{f}_{\boldsymbol{u}\mid t}\left(\cdot,t\right),\hat{f}_{\boldsymbol{v}\mid t}\left(\cdot,t\right)\right)=\int\hat{f}_{\boldsymbol{u}\mid t}\left(\boldsymbol{x}_{\boldsymbol{u}},t\right)\hat{f}_{\boldsymbol{v}\mid t}\left(\boldsymbol{x}_{\boldsymbol{v}},t\right)\mathrm{d}F\left(\boldsymbol{x}\right)=0,\quad\forall\boldsymbol{u}\ne\boldsymbol{v},\forall t.\label{eq:Cov Orthogonality}
\end{equation}

(c) Local variance decomposition: We have 
\begin{align*}
\mathbb{V}_{\boldsymbol{x}}\left(\hat{f}\left(\boldsymbol{x},t\right)\right) & =\mathbb{V}_{\boldsymbol{x}}\left(\sum_{\boldsymbol{u}\subseteq\mathcal{D}}\hat{f}_{\boldsymbol{u}\mid t}\left(\boldsymbol{x}_{\boldsymbol{u}},t\right)\right)\\
 & =\sum_{\boldsymbol{u},\boldsymbol{v}\subseteq\mathcal{D}}\mathrm{Cov}\left(\hat{f}_{\boldsymbol{u}\mid t}\left(\cdot,t\right),\hat{f}_{\boldsymbol{v}\mid t}\left(\cdot,t\right)\right)\\
 & =\sum_{\boldsymbol{u}\subseteq\mathcal{D}}\mathbb{V}_{\boldsymbol{x}}\left(\hat{f}_{\boldsymbol{u}\mid t}\left(\boldsymbol{x}_{\boldsymbol{u}},t\right)\right)+\sum_{\boldsymbol{u}\ne\boldsymbol{v}}\mathrm{Cov}\left(\hat{f}_{\boldsymbol{u}\mid t}\left(\cdot,t\right),\hat{f}_{\boldsymbol{v}\mid t}\left(\cdot,t\right)\right)\\
 & =\sum_{\boldsymbol{u}\subseteq\mathcal{D}}\mathbb{V}_{\boldsymbol{x}}\left(\hat{f}_{\boldsymbol{u}\mid t}\left(\boldsymbol{x}_{\boldsymbol{u}},t\right)\right),
\end{align*}
where the first equality follows from FOED, the second equality comes
from the definition of variance as the sum of covariances between
all pairs of effect functions, the third equality splits the total
variance into individual variances and cross-effect covariances, and
the last equality follows from the conditional orthogonality as shown
in Eq. \eqref{eq:Cov Orthogonality}. 

(d) Global variance decomposition: This result directly follows by
taking the expectation with respect to the local variance decomposition,
which finishes the proof. \end{proofTheorem}

\begin{proofTheorem} Utilizing the independence between the input
variable $\boldsymbol{x}$ and time $t$, along with the conditional
mean given by Theorem 2(b), we have 
\begin{align*}
V_{\boldsymbol{u}\mid t}\left(t\right) & =\mathbb{V}_{\boldsymbol{x}_{\boldsymbol{u}}}\left(\hat{f}_{\boldsymbol{u}\mid t}\left(\boldsymbol{x}_{\boldsymbol{u}},t\right)\right)\\
 & =\mathbb{V}_{\boldsymbol{x}_{\boldsymbol{u}}}\left(\left(\delta_{t}^{2}\boldsymbol{k}_{t}\odot\left(\bigodot_{i\in\boldsymbol{u}}\delta_{i}^{2}\boldsymbol{k}_{i}\right)\right)^{\top}\boldsymbol{\gamma}\right)\\
 & =\boldsymbol{\gamma}^{\top}\mathbb{V}_{\boldsymbol{x}_{\boldsymbol{u}}}\left(\left(\delta_{t}^{2}\boldsymbol{k}_{t}\odot\left(\bigodot_{i\in\boldsymbol{u}}\delta_{i}^{2}\boldsymbol{k}_{i}\right)\right)^{\top}\right)\boldsymbol{\gamma}\\
 & =\boldsymbol{\gamma}^{\top}\left(\delta_{t}^{4}\boldsymbol{k}_{t}\boldsymbol{k}_{t}^{\top}\right)\odot\left(\bigodot_{i\in\boldsymbol{u}}\mathbb{V}_{x_{i}}\left(\delta_{i}^{2}\boldsymbol{k}_{i}\right)\right)\boldsymbol{\gamma}\\
 & =\boldsymbol{\gamma}^{\top}\left(\delta_{t}^{4}\boldsymbol{k}_{t}\boldsymbol{k}_{t}^{\top}\right)\odot\left(\bigodot_{i\in\boldsymbol{u}}\delta_{i}^{4}\mathbb{E}_{x_{i}}\left[\boldsymbol{k}_{i}\boldsymbol{k}_{i}^{\top}\right]\right)\boldsymbol{\gamma},
\end{align*}
where the third equality uses the property $\mathbb{V}\left(\boldsymbol{XB}\right)=\boldsymbol{B}^{\top}\mathbb{V}\left(\boldsymbol{X}\right)\boldsymbol{B}$,
the forth equality leverages the independence between the input variable
$\boldsymbol{x}$ and time $t$ and the property of Hadamard product,
and the last equality follows from the identity $\mathbb{V}\left(\boldsymbol{X}\right)=\mathbb{E}\left[\boldsymbol{XX}^{\top}\right]-\mathbb{E}\left[\boldsymbol{X}\right]\mathbb{E}\left[\boldsymbol{X}\right]^{\top}$
and the property $\mathbb{E}_{x_{i}}\left[\tilde{k}_{i}\left(x_{i},x_{i}^{\prime}\right)\right]=0$.
Taking the expectation of $V_{\boldsymbol{u}\mid t}\left(t\right)$
with respect to time $t$ yields
\begin{align*}
V_{\boldsymbol{u}} & =\mathbb{E}_{t}\left[\boldsymbol{\gamma}^{\top}\left(\delta_{t}^{4}\boldsymbol{k}_{t}\boldsymbol{k}_{t}^{\top}\right)\odot\bigodot_{i\in\boldsymbol{u}}\delta_{i}^{4}\mathbb{E}_{x_{i}}\left[\boldsymbol{k}_{i}\boldsymbol{k}_{i}^{\top}\right]\boldsymbol{\gamma}\right]\\
 & =\boldsymbol{\gamma}^{\top}\left(\delta_{t}^{4}\mathbb{E}_{t}\left[\boldsymbol{k}_{t}\boldsymbol{k}_{t}^{\top}\right]\odot\left(\bigodot_{i\in\boldsymbol{u}}\delta_{i}^{4}\mathbb{E}_{x_{i}}\left[\boldsymbol{k}_{i}\boldsymbol{k}_{i}^{\top}\right]\right)\right)\boldsymbol{\gamma},
\end{align*}
which finishes the proof. \end{proofTheorem}
\end{document}